\documentclass[10pt,letterpaper,journal]{IEEEtran}
\usepackage{algorithmic}
\usepackage{algorithm}
\usepackage{array}
\usepackage[caption=false,font=normalsize,labelfont=sf,textfont=sf]{subfig}
\usepackage{textcomp}
\usepackage{stfloats}
\usepackage{url}
\usepackage{verbatim}
\usepackage{graphicx}
\usepackage{cite}
\usepackage{amsmath,amssymb,amsfonts,amsthm,mathtools}
\usepackage{xcolor}
\usepackage{blindtext}
\usepackage{lipsum}
\usepackage{subfig}
\usepackage{cuted}
\usepackage{physics}
\usepackage[long]{optidef}
\usepackage{multirow}
\usepackage{steinmetz}
\usepackage{bigints}

\usepackage{physics}
\def\BibTeX{{\rm B\kern-.05em{\sc i\kern-.025em b}\kern-.08em
    T\kern-.1667em\lower.7ex\hbox{E}\kern-.125emX}}

\usepackage{eso-pic}
\usepackage{balance}
\begin{document}
\title{Optimal Illumination via Joint Movement and Phase Optimization for Movable Antenna-RIS Configuration}

\author{Yan~Zhang,~\IEEEmembership{Student~Member,~IEEE,}
Nicola~Marchetti,~\IEEEmembership{Senior Member,~IEEE,}
and Indrakshi~Dey,~\IEEEmembership{Senior Member,~IEEE}
\thanks{Y.~Zhang and N.~Marchetti are with the the Department of Electrical and Electronic Engineering, Trinity College Dublin, Dublin, Ireland. (Email: zhangy42@tcd.ie, nicola.marchetti@tcd.ie)}
\thanks{I. Dey is with the Walton Institute, South East Technological University, Waterford, Ireland. (Email: indrakshi.dey@waltoninstitute.ie)}

\thanks{This work was supported in part by the Chinese Scholarship Council. This work was also supported in part by Taighde Eireann - Research Ireland under Grant 13/RC/2077\_P2, by the EU MSCA Project “COALESCE” under Grant Number 101130739, and by US-Ireland R\&D
Partnership Programme RI-SFI-23/US/3924.}}
\markboth{Journal of \LaTeX\ Class Files,~Vol.~18, No.~9, September~2020}%
{Joint Movement and Phase Optimization for Optimal Illumination for Movable Antenna-RIS Configuration}
\maketitle
\begin{abstract}
Reconfigurable intelligent surfaces (RIS) enable programmable control of wireless propagation but remain vulnerable to persistent deep fades in static deployments. This paper introduces a Movable Antenna-enhanced RIS (MA-RIS) architecture where antenna elements physically reposition to sample independent spatial channels, enabling mobility-induced diversity. We model antenna motion using a Stochastic Differential Equation (SDE) framework capturing controlled drift and environmental diffusion. Itô calculus-based analysis characterizes steady-state antenna distributions, spatial decorrelation, and outage probability, revealing fundamental trade-offs between control strength and mobility randomness. To maximize long-term SNR while accounting for control overhead, we propose an overhead-aware Two-timescale framework separating slow antenna trajectory control from fast phase adaptation. The stochastic optimal control problem is solved via predictive approximation of the Hamilton-Jacobi-Bellman (HJB) formulation, enabling real-time implementation. Simulations validate theoretical predictions: the Two-timescale strategy achieves up to 36 dB steady-state SNR with remarkable stability, outperforming position-only control by up to 15 dB and uncontrolled baselines by over 30 dB. Despite experiencing a lower SNR than Active RIS, the proposed approach delivers up to 16 times higher energy efficiency (EE) across varying system scales, establishing a new paradigm of mobility-enabled channel adaptation for resilient wireless systems.
\end{abstract}

\begin{IEEEkeywords}
RIS, Movable antenna, Stochastic optimal control, Two-timescale optimization, Energy Efficiency, Outage.
\end{IEEEkeywords}

\section{Introduction}
\subsection{Background and Related Work}
The proliferation of wireless services and Internet-of-Things (IoT) devices has intensified the demand for spectrum- and energy-efficient communication solutions \cite{shafi20175g}. Traditional approaches to improving wireless link quality rely primarily on sophisticated signal processing techniques, advanced modulation schemes, and multi-antenna technologies \cite{wang2024tutorial}. However, these methods treat the wireless channel as an uncontrollable black box, passively adapting to random propagation conditions rather than actively shaping the electromagnetic environment \cite{bilotti2024reconfigurable}.

RIS have recently emerged as a transformative technology that challenges this paradigm by enabling programmable control over wireless propagation \cite{wu2019towards}. An RIS consists of a planar array of passive or semi-passive elements that can independently adjust the phase, amplitude, and polarization of incident electromagnetic waves, effectively transforming the wireless channel into a controllable resource \cite{wu2019intelligent}. By intelligently configuring these elements, RIS can create favorable propagation paths, suppress interference, and extend coverage to shadowed regions \cite{huang2019reconfigurable, zhang2020capacity}. The passive nature of RIS elements also offers significant advantages in EE and hardware cost compared to conventional active relay systems \cite{lu2023secrecy, qiao2023joint, wang2023joint}.

Despite these promising features, RIS deployments face a fundamental limitation: when the channel between the RIS and the receiver experiences a deep fade due to destructive multipath interference or severe shadowing, the received signal power can drop dramatically regardless of phase optimization \cite{yang2020intelligent,yu2020robust}. Unlike spatial diversity techniques in conventional multiple-input multiple-output (MIMO) systems where multiple spatially separated antennas provide independent fading realizations \cite{tse2005fundamentals}, a static RIS cannot escape persistently poor channel conditions. This susceptibility to static deep fades critically undermines the reliability of RIS-assisted communications, particularly for mission-critical applications requiring ultra-reliable low-latency communications (URLLC) \cite{bennis2018ultrareliable}.

Extensive research has investigated various aspects of RIS-assisted communication systems. Early works focused on channel modeling and capacity analysis, establishing fundamental performance limits of RIS-aided links \cite{wu2019beamforming,guo2020weighted,yu2019miso}. Subsequent studies addressed the optimization of RIS phase shifts to maximize signal-to-noise ratio (SNR) or minimize bit error rate, typically formulating non-convex optimization problems solved via alternating optimization, semidefinite relaxation, or manifold optimization techniques \cite{huang2021multi,zheng2019intelligent,jensen2020optimal}. Multi-user scenarios have been explored, addressing fairness, sum-rate maximization, and EE through joint active and passive beamforming \cite{peng2022deep,shvetsov2023federated}.
These approaches still result in static configurations that cannot adapt to time-varying channel conditions. The concept of reconfigurability beyond phase adjustment---such as amplitude control or switching networks---has been explored \cite{zhang2022active,fu2021reconfigurable}, but these modifications cannot overcome deep fades arising from fundamental geometric limitations of fixed positions.

The emerging paradigm of Movable Antenna (MA) systems offers a potential solution by enabling physical repositioning of antenna elements to access different spatial channel realizations \cite{zhu2023movableop,new2024tutorial,zhu2025tutorial}. Initial works on MA-MIMO demonstrated that modest antenna movements on the order of half-wavelength can significantly improve channel conditions by avoiding spatial nulls \cite{zhu2023movable}. More recently, the integration of movable antennas into RIS architectures has been proposed, where elements within the surface adjust their positions to optimize channel quality \cite{wang2024movable}. Zhang et al.~\cite{zhang2024ris} evaluated MA-enabled RIS (MA-RIS) against fixed-position antenna RIS, demonstrating significant performance gains in outage probability and SNR by analyzing the geometry impact of element positions in 1-Dimension and 2 Dimension (2D) array configurations. However, their analysis assumes idealized antenna placements and does not model continuous motion dynamics, formulate position optimization, or account for the joint optimization of antenna trajectories and phase shifts under realistic physical perturbations. More broadly, these works typically assume deterministic movement with perfect position control, treating antenna repositioning as a one-shot geometric problem rather than a continuous stochastic process subject to thermal fluctuations in Micro Electro Mechanical Systems (MEMS) actuators, structural vibrations transmitted through the mounting platform, and wind-induced perturbations for outdoor deployments. 

This paper addresses these limitations through three key contributions. First, current models lack a physically grounded treatment of antenna dynamics: practical MEMS actuators exhibit random perturbations that are naturally captured by SDE, yet no existing work models movable antenna motion in this framework. Second, the joint optimization of antenna trajectories and RIS phase shifts has not been addressed in a principled manner---existing approaches optimize positions and phases independently or through heuristic alternation, without accounting for the coupling between spatial dynamics and phase coherence. Third, and perhaps most critically, the overhead cost of Channel State Information (CSI) acquisition scales with the number of effective antennas: as position control concentrates antennas near the illumination center and increases effective number, the pilot overhead grows correspondingly, creating a fundamental tension between beamforming gain and spectral efficiency that existing works ignore. 
This motivates a Two-timescale optimization framework in which slow-timescale position control and fast-timescale phase alignment are jointly designed to balance overhead cost against robust SNR performance---precisely the approach developed in this paper.

\subsection{Contributions}
This paper addresses the aforementioned challenges by developing a unified theoretical and algorithmic framework for movable antenna–enhanced RIS systems that integrates stochastic mobility modeling, performance analysis, and overhead-aware joint optimization. Our main contributions are summarized as follows:
\subsubsection{Stochastic Mobility Modeling for MA-RIS}
We develop an SDE framework to model MA element dynamics under joint deterministic control (drift toward optimal regions) and environmental randomness (Brownian diffusion). Using Itô calculus, we derive closed-form antenna position evolution and establish steady-state statistical properties, revealing a fundamental tradeoff between drift control strength $V_{\text{drift}}$ and diffusion coefficient $\sigma$ in determining average antenna distance from the illumination center. We further derive explicit expressions for the effective antenna count $N_{\mathrm{eff}}$ and characterize how stochastic mobility impacts coherent combining gain, providing tractable analytical tools for MA-RIS system design.

\subsubsection{Outage Probability Analysis}
We develop an analytical outage probability framework using a moment-generating function approach combined with Gaussian approximation techniques. The resulting closed-form expressions quantify reliability as a function of drift velocity $V_{\mathrm{drift}}$, diffusion coefficient $\sigma$, array size $N$, and SNR threshold $\gamma_{\mathrm{th}}$, providing explicit design guidelines for satisfying probabilistic performance constraints in MA-RIS systems.

\subsubsection{Overhead-Aware Two-timescale Joint Movement–Phase Optimization}
We propose a novel overhead-aware Two-timescale joint optimization framework that separates slow-timescale antenna trajectory control from fast-timescale RIS phase adaptation. The problem is formulated as a stochastic optimal control task that maximizes long-term effective SNR, while explicitly accounting for mobility and channel estimation overhead. To address the computational complexity of the continuous-state HJB formulation, we develop a predictive discrete dynamic programming approximation that enables real-time implementation. The proposed Two-timescale strategy achieves near-optimal performance with substantially reduced control overhead compared to fully joint per-slot optimization.

\textbf{Notation:} Throughout this paper, boldface lowercase and uppercase letters denote vectors and matrices, respectively. The operators $(\cdot)^T$, $(\cdot)^*$, $\|\cdot\|$, and $|\cdot|$ represent transpose, conjugate, Euclidean norm, and absolute value/cardinality. $\mathrm{E}[\cdot]$, $\text{Var}[\cdot]$, and $\text{tr}(\cdot)$ denote expectation, variance, and trace operators. $\mathbb{R}$ and $\mathbb{C}$ denote real and complex number fields, with $\mathbb{R}^+$ denoting positive real numbers. $\mathbb{R}^{m \times n}$ and $\mathbb{C}^{m \times n}$ denote the space of $m \times n$ real and complex matrices, respectively. $\mathcal{J}_0(\cdot)$ is the zeroth-order Bessel function of the first kind. The notation $\mathcal{N}(\mu, \Sigma)$ represents a Gaussian distribution with mean $\mu$ and covariance $\Sigma$, and $\mathcal{CN}(\mu, \Sigma)$ represents a complex Gaussian distribution. $Q(\cdot)$ denotes the standard Q-function, and $\text{erf}(\cdot)$ is the error function. $\text{diag}\{\cdot\}$ denotes a diagonal matrix. $j = \sqrt{-1}$ is the imaginary unit.

\section{System Model}

\begin{figure}
    \centering
    \makebox[\linewidth][c]{%
        \includegraphics[width=0.7\linewidth, trim=0 5 80 40, clip]{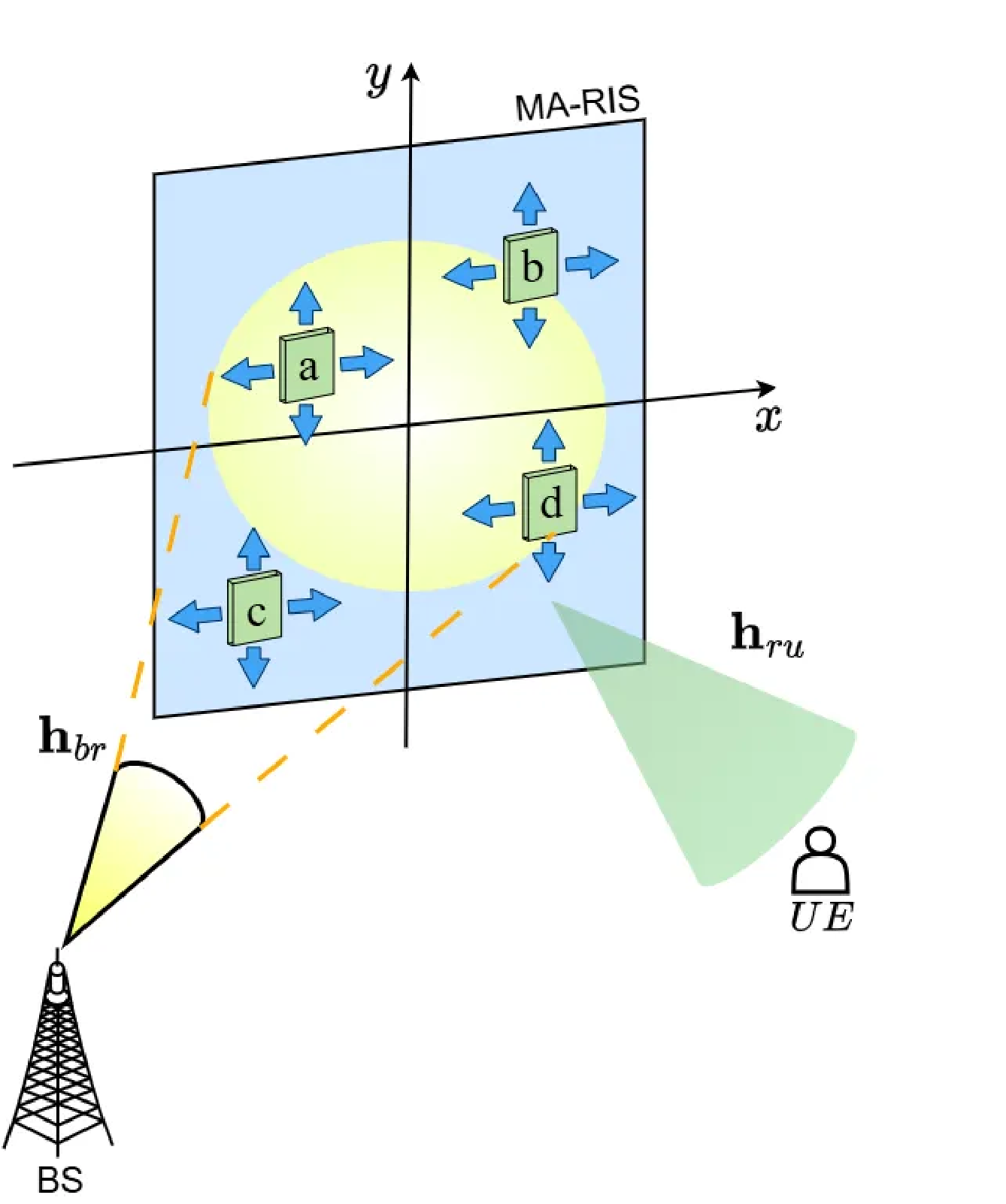}}
\caption{MA-RIS assisted wireless communication system architecture. A base station (BS) transmits an incident beam with finite beamwidth (yellow illuminated region) toward the MA-RIS panel deployed in the $xy$-plane. The panel contains MAs that can physically relocate within the panel region (blue area) to sample different spatial channel realizations. Blue arrows indicate the 2D mobility capability of each antenna element. The orange dashed lines trace the beam's propagation cone from the BS to the MA-RIS panel, visualizing the spatial coverage of the incident wave.} 
\label{fig:system_model} 
\vspace{-0.5mm}
\end{figure}
\subsection{MA-RIS System Architecture and Mobility Framework}

The MA-RIS system consists of $N$ antenna elements deployed on a planar panel in the $xy$-coordinate plane (illustrated as the blue rectangular region in Fig.~\ref{fig:system_model}). Unlike conventional static RIS, each antenna element can physically relocate within a confined panel region to adaptively sample different spatial channels. The center of the MA-RIS panel is positioned at $\mathbf{r}_c = [x_r, y_r, z_r]^T$ in the 3D coordinate system, and the time-varying position of the $n$-th element is denoted as $\mathbf{X}_n(t) = [x^r_n(t), y^r_n(t)]^T \in C_r$ for $n \in \mathcal{N} = \{1, \ldots, N\}$, where $C_r \subset \mathbb{R}^2$ defines the bounded 2D movement region on the panel plane, designed to prevent collisions while allowing sufficient spatial diversity.

Each MA element follows stochastic movement governed by controlled drift toward optimal positions (seeking the illumination center) and environmental diffusion representing mechanical noise and perturbations. Four representative elements labeled as a, b, c, and d in Fig.~\ref{fig:system_model} demonstrate different positioning scenarios:
\begin{itemize}
\item \textbf{Element a:} Optimally positioned at the illumination center, receiving maximum incident power and contributing most effectively to the reflected signal with high effectiveness.
\item \textbf{Elements b and d:} Positioned at the edge of the illumination region, partially illuminated with reduced incident power density and intermediate effectiveness.
\item \textbf{Element c:} Located outside the illumination zone, receiving minimal incident power and contributing negligibly to beamforming.
\end{itemize}

Unlike idealized models assuming instantaneous repositioning, practical MA elements experience physical constraints including velocity limits, actuation noise, mechanical perturbations, and environmental disturbances. These considerations motivate our stochastic mobility model accounting for both controlled movement and random perturbations. This architecture enables MA-RIS to adaptively reconfigure spatial structure and electromagnetic response, providing 2D optimization unavailable to conventional static RIS.

\subsection{Time-Varying Cascaded Channel Model}

The end-to-end channel from Tx to Rx via the MA-RIS exhibits time-varying characteristics due to element mobility. We adopt a line-of-sight (LoS) dominated propagation model, appropriate for high-frequency RIS deployments. The Tx and Rx are located at fixed positions $\mathbf{b} = [x_b, y_b, z_b]^T \in \mathbb{R}^{3 \times 1}$ and $\mathbf{u} = [x_u, y_u, z_u]^T \in \mathbb{R}^{3 \times 1}$, respectively. For the single-antenna transmitter at position $\mathbf{b}$, the array response vector relative to the reference point $\mathbf{b}_0 = [0, 0, z_b]^T \in \mathbb{R}^{3 \times 1}$ at the MA-RIS panel center is $s(\mathbf{b}) = \exp\left(j\frac{2\pi}{\lambda}\rho(\mathbf{b})\right) $, where $\rho(\mathbf{b}) = \|\mathbf{b} - \mathbf{b}_0\| \in \mathbb{R}$ is the distance from Tx to the reference point, and $\lambda \in \mathbb{R}$ is the carrier wavelength.

\textbf{Tx-to-MA-RIS Link:} The time-varying channel vector from Tx to the MA-RIS is
\begin{equation}
\mathbf{h}_{br}(t) = \alpha_n(t) \mathbf{G}^r(\mathbf{X}(t))^H s(\mathbf{b}) \in \mathbb{C}^{1 \times N}
\label{eq:hbr}
\end{equation}
where $\mathbf{G}^r(\mathbf{X}(t)) \in \mathbb{C}^{N \times 1}$ is the time-varying MA-RIS array response vector for the Tx-to-RIS link, given by $\mathbf{G}^r(\mathbf{X}(t)) = [g^r(\mathbf{X}_1(t)), g^r(\mathbf{X}_2(t)), \ldots, g^r(\mathbf{X}_N(t))]^T$, with each element $g^r(\mathbf{X}_n(t)) = \exp\left(j\frac{2\pi}{\lambda}\rho^r(\mathbf{X}_n(t))\right) \in \mathbb{C}$. Here, $\rho^r(\mathbf{X}_n(t)) = \|\mathbf{b} - \mathbf{r}_n(t)\| - \|\mathbf{b} - \mathbf{r}_0\| \in \mathbb{R}$ represents the path difference from Tx to the $n$-th element at position $\mathbf{r}_n(t) = [x^r_n(t), y^r_n(t), z^r_n(t)]^T \in \mathbb{R}^{3 \times 1}$ relative to the reference point, where $\mathbf{X}_n(t) = [x^r_n(t), y^r_n(t)]^T \in \mathbb{R}^{2 \times 1}$ denotes the 2D projection of the antenna position on the panel plane. 

The small-scale fading coefficient $\alpha_n(t)\in\mathbb{C}$ for element $n$ is modeled as a LoS-dominated \emph{Rician} random process: 
\begin{equation}
\alpha_n(t) = \sqrt{\frac{K_{br}}{K_{br}+1}}\,\bar{\alpha}_n +
\sqrt{\frac{1}{K_{br}+1}}\,\tilde{\alpha}_n\!\big(\mathbf{X}_n(t)\big)
\end{equation}
where $K_{br}\ge 0$ is the Rician $K$-factor, $\bar{\alpha}_n\in\mathbb{C}$ denotes the deterministic LoS component, and $\tilde{\alpha}_n(\cdot)\sim\mathcal{CN}(0,\sigma_\alpha^2)$ is the diffuse (NLoS) component. Under isotropic scattering, the diffuse component exhibits Jakes spatial correlation as:
\begin{equation}
  \mathrm{E}\!\left[\tilde{\alpha}_n(\mathbf{X})\,\tilde{\alpha}_n^*(\mathbf{X}+\Delta \mathbf r)\right] =\sigma_\alpha^2\,J_0\!\left(\frac{2\pi}{\lambda}\|\Delta \mathbf r\|\right)  
  \label{eq:alpha_corr}
\end{equation}
\textbf{MA-RIS-to-Rx Link:} The time-varying channel vector from MA-RIS to Rx is
\begin{equation}
\mathbf{h}_{ru}(t)= f(\mathbf{u})\,
\big[\, \beta_n(t)\, g^t(\mathbf{X}_n(t)) \,\big]_{n=1}^N
\in \mathbb{C}^{1\times N},
\label{eq:hru}
\end{equation}
where $f(\mathbf{u})$ denotes the large-scale attenuation (pathloss and shadowing) toward the receiver at location $\mathbf{u}$, $g^t(\mathbf{X}_n(t))=\exp\!\left(j\frac{2\pi}{\lambda}\rho^t(\mathbf{X}_n(t))\right)$ is the deterministic LoS array response, $\rho^t(\mathbf{X}_n(t)) =\|\mathbf{r}_n(t)-\mathbf{u}\|-\|\mathbf{r}_0-\mathbf{u}\|$ represents the path difference from $n$-th element to the receiver at position $\mathbf{u}$, and $\beta_n(t)$ is the small-scale fading coefficient of the MA-RIS$\to$Rx hop for element $n$. Similarly, we model $\beta_n(t)$ as a LoS-dominated \emph{Rician} random process: $\beta_n(t) = \sqrt{\frac{K_{ru}}{K_{ru}+1}}\,\bar{\beta}_n + \sqrt{\frac{1}{K_{ru}+1}}\,\tilde{\beta}_n\!\big(\mathbf{X}_n(t)\big)$, where $K_{ru}\ge 0$ is the Rician $K$-factor, $\bar{\beta}_n$ denotes the deterministic LoS component, and $\tilde{\beta}_n(\cdot)\sim\mathcal{CN}(0,\sigma_\beta^2)$ is the NLoS component, whose spatial correlation follows $\mathrm{E}\!\left[\tilde{\beta}_n(\mathbf{X})\,\tilde{\beta}_n^*(\mathbf{X}+\Delta \mathbf r)\right] =\sigma_\beta^2\,J_0\!\left(\frac{2\pi}{\lambda}\|\Delta \mathbf r\|\right)$.

For a unit-power transmit symbol $s$ with $\mathrm{E}[|s|^2]=1$, the received signal is
\begin{equation}
y(t)=\sqrt{P_t}\,\mathbf{h}_{ru}(t)\boldsymbol{\Theta}(t)\mathbf{h}_{br}(t)\,s+n_0,
\label{eq:rx_signal}
\end{equation}
where $P_t$ is the transmit power, and $n_0\sim\mathcal{CN}(0,\sigma_n^2)$ is the receiver noise. The MA-RIS applied time-varying phase shifts is $\boldsymbol{\Theta}(t)=\mathrm{diag}\!\left\{e^{j\theta_1(t)},\ldots,e^{j\theta_N(t)}\right\}$, where $\theta_n(t)\in[0,2\pi)$. Substituting the channel expressions  (\ref{eq:hbr}) and (\ref{eq:hru}) yields
\begin{equation}
y(t)=\sqrt{P_t}\sum_{n=1}^{N_{\mathrm{eff}}(t)}
\eta_n(t)\,
\exp\!\left(j\Big(\theta_n(t)-\tfrac{2\pi}{\lambda}\rho_n(t)\Big)\right)s+n_0,
\label{eq:rx_signal_sum}
\end{equation}
where $N_{\mathrm{eff}}(t)$ is the instantaneous number of effective contributing MA elements, and $\eta_n(t)=\alpha_n(t)\beta_n(t)$ denotes the complex channel gains. $\rho_n(t)$ denotes the total path-difference term from Tx through the $n$-th MA element to Rx. The geometric terms $\rho^t(\mathbf{X}_n(t))$ and $\rho_n(t)$ capture deterministic LoS phase evolution due to movement, while the small-scale fading that creates deep fades is represented by the diffuse components $\tilde{\alpha}_n(\cdot)$ and $\tilde{\beta}_n(\cdot)$. The Jakes correlation model in \eqref{eq:alpha_corr} therefore applies to the NLoS part, and the overall model is LoS-dominated when $K_{br}$ and $K_{ru}$ are large.

\section{Fading Mitigation via Movement}
In a static RIS, the received power may sharply decrease if the channel between the RIS and the receiver experiences a deep fade. However, movement enables sampling of multiple independent spatial channels over time, similar to spatial diversity gains. Mathematically, assuming small displacements over time $t$, the received channel coefficients at the receiver for the $n$th antenna can be approximated as,
\begin{align}
    h_n(t) = \alpha_n \beta_n \,e^{j (\theta_n(t) - 2\pi\rho_n(t)/\lambda)}
\end{align}
where $\alpha_n$ and $\beta_n$ are fading coefficients from transmitter to RIS and RIS to receiver respectively. $\rho_n(t) =\rho(\mathbf{b})+\rho^r(\mathbf{X}_n(t)) + \rho^t(\mathbf{X}_n(t)) $ denotes the end-to-end path difference from Tx to Rx via the $n$-th MA element. Since $\rho(\mathbf{b})$ and $\rho(\mathbf{u})$ are fixed, movement causes $\rho_n(t)$ to vary through the position-dependent terms $\rho^t(\mathbf{X}_n(t))$ and $\rho^r(\mathbf{X}_n(t)),$ continuously de-correlating the fading and allowing the system to escape static deep fades. Now the spatial correlation $R(\Delta r)$ between two MA positions separated by a displacement $\Delta r$ can be modeled as,
\begin{align}
    R(\Delta r) = \mathcal{J}_0(2\pi \Delta r/\lambda)
    \label{eq:spa_corr}
\end{align}
where $\mathcal{J}_0$ is the zeroth-order Bessel function and $\lambda$ is the wavelength of operation. With simple mathematical calculations, we can conclude that if $\Delta r >> 0.4 \lambda$, spatial fading becomes effectively uncorrelated. Thus, by moving a distance of at least 0.4$\lambda$, the MA-RIS can refresh the link and avoid deep persistent fades. The correlation model in (\ref{eq:alpha_corr}) applies to the \emph{diffuse (NLoS) component} of the Rician fading field, like, $\tilde{\alpha}_n(X)$ and $\tilde{\beta}_n(X)$ under isotropic scattering.
The LoS terms are deterministic and do not decorrelate with displacement in the same manner. Therefore, moving an MA element by $\Delta r\gtrsim 0.4\lambda$ primarily \emph{refreshes the diffuse multipath contribution}, yielding an approximately independent small-scale fading realization, which is the mechanism by which MA-RIS escapes persistent deep fades.

While movement brings diversity, it also introduces new impairments, in particular Doppler effect. The 
MA movement at velocity $v$ results in Doppler frequency $f_D = (v/\lambda) \cos(\theta)$, which results in a time-varying channel with a coherence time of $T_c \approx 1/2\pi f_D$. With a carrier frequency of 4.25 GHz, antenna element movement speed of 0.1 m/s and an aligned movement of $\theta = 0^o$, $f_D \approx 1.43$ GHz, and $T_c \approx 0.11$ seconds. Since the cascaded MA-RIS channel must be estimated within each coherence interval, it is generally impractical to continuously move antennas while simultaneously performing accurate channel estimation and coherent reflection. This motivates a Two-timescale \emph{move--estimate--reflect} protocol (defined below) that explicitly accounts for CSI overhead and estimation errors. Since the antenna panel movement is relatively slow, the Doppler spread is minimal with a channel coherence time that is large compared to typical symbol durations. However, with a higher movement velocity, like drone-mounted RIS panels for example, the Doppler shift will be non-negligible, necessitating adaptive channel estimation techniques, an important generalization that we will explore in our future work.

There will also be an impact on the effective number of elements, $N_{\text{eff}}$. This is because the illuminated area depends on the instantaneous position of MA-RIS elements. With higher movement velocities for the elements, the position of each element changes quickly. 
Therefore, within a given coherence time, the elements can enter or leave the illuminated area frequently, causing fast changes in $N_{\text{eff}}$ and leading to fluctuating SNR at the receiver. The average effective number of elements over a duration $T$ can be estimated as,
\begin{align}
    \bar{N}_{\text{eff}}(v) = \frac{1}{T}\int_0^T N_{\text{eff}}(t,v)\mathrm{d}t
\end{align}
where $N_{\text{eff}}(t,v)$ is the instantaneous effective number of illuminated elements depending on velocity $v$ and time $t$ and $N_{\text{eff}}(t,v) = N\,e^{-(vt)^2}$, which follows a Gaussian illumination model. This form is commonly adopted when the spatial illumination profile is Gaussian and the element displacement is linear in time ($x=vt$), yielding a Gaussian temporal weighting. Such modeling is standard in mobility-induced fading analyses and beam illumination studies. Then,
\begin{align}
    \bar{N}_{\text{eff}}(v) = \frac{N}{T}\int_0^T \,e^{-(vt)^2}\mathrm{d}t.
\end{align}
Let $u = vt, t = u/v, \mathrm{d}t = \mathrm{d}u/v$, then the limits change from $t = 0 \to T$ to $u = 0 \to vT$. So we get, 
\begin{align}
    \bar{N}_{\text{eff}}(v) &= \frac{N}{T}\int_0^T \,e^{-(vt)^2}\mathrm{d}t \nonumber\\
    &= \frac{N}{T} \times \frac{1}{v}\int_0^{vT} \,e^{-u^2}\mathrm{d}u = \frac{N\sqrt{\pi}}{2Tv}\,\operatorname{erf}(vT)\nonumber\\
\end{align}
where $\,erf(\cdot)$ is the error function. For small $v$ (i.e., $vT \ll 1$), using the first-order Taylor expansion $\operatorname{erf}(x) \approx \frac{2}{\sqrt{\pi}}x$, we obtain $\operatorname{erf}(vT) \approx \frac{2}{\sqrt{\pi}}vT$, hence $\tilde{N}_{\text{eff}} \approx N$ (full activation). For large $v$ (i.e., $vT \gg 1$), $\operatorname{erf}(vT) \to 1$, and thus $\tilde{N}_{\text{eff}} \approx \frac{N\sqrt{\pi}}{2Tv}$, which decays proportionally to $(1/v)$.

\subsection{Mobility-Aware MA-RIS System Modeling}

\subsubsection{Stochastic Movement Model}
We model the mobility of each MA element as a continuous-time stochastic process using SDE framework:
\begin{align}\label{eq:sde}
    \mathrm{d}\mathbf{X}_n(t) = \mathbf{V}_n(t)\mathrm{d}t + \sigma \mathrm{d}\mathbf{W}_n(t)
\end{align}
where $\mathbf{X}_n(t) \in \mathbb{R}^2$ is the position vector of antenna $n$ at time $t$, $\mathbf{V}_n(t)$ is the deterministic drift velocity (control input), $\sigma \geq 0$ is the mobility diffusion coefficient, and $\mathbf{W}_n(t)$ is a standard 2D Wiener process. The Wiener process further satisfies,
\begin{align}
    \mathbf{W}_n(0) = \mathbf{0}; \quad \mathbf{W}_n(t) \sim \mathcal{N}(\mathbf{0}, t\mathbf{I}_2)
\end{align}
where $\mathbf{I}_2$ is the 2D Identity matrix and $\mathbf{W}_n(t) - \mathbf{W}_n(s) \approx \mathcal{N}(0, (t-s)\mathbf{I}_2)$ for $t > s$ represents independent stationary increments between time $s$ and $t$, and $\mathcal{N}(\cdot)$ represents Normal distribution. The above mathematically presents how the MA position evolves according to both controlled movement and random environmental or mechanical perturbations. Although the MA dynamics are modeled using an SDE, the antenna trajectories are restricted to the feasible panel region $\mathcal{C}_r$. Accordingly, the diffusion term $\sigma \mathrm{d}\mathbf{W}_n(t)$ models only local random perturbations due to actuator imperfections, platform vibration, or wind-induced micro-motion, rather than unrestricted large-scale displacement. To enforce the deterministic panel geometry, the process is understood as a constrained diffusion on $\mathcal{C}_r$; i.e., whenever a tentative update reaches the boundary $\partial\mathcal{C}_r$, the outward component is reflected (or equivalently projected) so that $\mathbf{X}_n(t) \in \mathcal{C}_r$ for all $t$. Thus, the RIS surface boundaries remain hard physical constraints, while the diffusion term captures residual uncertainty within those constraints.

Next, we derive the explicit solution to the SDE for each antenna trajectory as,
\begin{align}
    \mathbf{X}_n(t) = \mathbf{X}_n(0) + \int_0^t\mathbf{V}_n(s)\mathrm{d}s + \sigma \mathbf{W}_n(t).
\end{align}
To analyze the evolution of antenna positions, we first consider a simplified scenario where the drift velocity is constant in direction and magnitude, $\mathbf{V}_n(t) = \mathbf{V}_0$ (constant vector). In this case, the solution simplifies to:
\begin{align}\label{e19}
    \mathbf{X}_n(t) = \mathbf{X}_n(0) + \mathbf{V}_0 t + \sigma \mathbf{W}_n(t),
\end{align}
which allows us to understand both deterministic movement (drift) and random dispersion (diffusion) quantitatively. With this explicit trajectory of antenna positions, we can now analyze derived quantities, like distance to a target (eg. center of illumination) to conclude on how movement of the MAs will affect the overall system performance. 
Using It$\hat{\text{o}}$ calculus, we next study the evolution of the squared distance of each antenna from a reference point. Applying It$\hat{\text{o}}$'s lemma \cite{oksendal2003stochastic} for functions of stochastic processes, the square distance to the illumination center evolves as,
\begin{align}
\hspace{-0.3cm} 
    \mathrm{d}||\mathbf{X}_n(t)||^2 = 2\mathbf{X}_n(t)^T\mathrm{d}\mathbf{X}_n(t) + \,\mathrm{tr}(\mathrm{d}\mathbf{X}_n(t)\mathrm{d}\mathbf{X}_n(t)^T)
    \label{e20}
\end{align}
where $T$ represents Transpose, $\mathbf{X}_n(t)^T\mathrm{d}\mathbf{X}_n(t)$ is a scalar dot product. Substituting (\ref{eq:sde}) in (\ref{e20}), we get,
\vspace{-2mm}
\begin{align}
\hspace{-0.3cm} 
    \mathrm{d}||\mathbf{X}_n(t)||^2 &= 2\mathbf{X}_n(t)^T\mathbf{V}_n(t)\mathrm{d}t + 2\sigma \mathbf{X}_n(t)^T\mathrm{d}\mathbf{W}_n(t) \nonumber\\
    &+ 2 \sigma^2 \mathrm{d}t.
    \label{e21}
\end{align}
Since the stochastic term $\mathbf{X}_n(t)^T\mathrm{d}\mathbf{W}_n(t)$ is a martingale increment, its expectation is zero; $\mathrm{E}[\mathbf{X}_n(t)^T\mathrm{d}\mathbf{W}_n(t)] = 0$, and therefore applying expectation to (\ref{e21}):
\begin{align}\label{e22}
    \frac{\mathrm{d}}{\mathrm{d}t}\mathrm{E}\big[||\mathbf{X}_n(t)||^2\big] = 2\mathrm{E}[\mathbf{X}_n(t)^T \mathbf{V}_n(t)] + 2\sigma^2.
\end{align}
The expected distance from the center impacts signal strength and activation probability. Next, we analyze the long-term behavior of the antenna position or evolution of expected distance, under the control policy that always moves them towards the center. If the drift is always towards the illumination center, i.e., $ \mathbf{V}_n(t) = - V_{\text{drift}}\frac{\mathbf{X}_n(t)}{||\mathbf{X}_n(t)||}$, it follows that $\mathrm{E}[\mathbf{X}_n(t)^T\mathbf{V}_n(t)] = - V_{\text{drift}}\mathrm{E}\big[||\mathbf{X}_n(t)||\big]$, and the steady-state equation then becomes,
\begin{align}\label{e25}
    V_{\text{drift}}\mathrm{E}\big[||\mathbf{X}_n(t)||\big] = \sigma^2
\end{align}
leading to, $\mathrm{E}\big[||\mathbf{X}_n(t)||\big] = \sigma^2/V_{\text{drift}}$. This equilibrium distance explicitly quantifies the trade-off between control and randomness: stronger drift velocity $V_{\text{drift}}$ pulls antennas closer to the illumination center, while larger diffusion coefficient $\sigma$ increases thermal fluctuations that scatter antennas away from the  illumination center. The equilibrium represents the steady-state balance where the deterministic control force exactly compensates for the stochastic spreading due to Brownian motion. The closer the antennas are to the center on average, the more effective they are, contributing to a higher effective antenna count and higher SNR over the communication link. This sets the stage for analyzing outage probability which depends on how link SNR changes over time.

\subsubsection{Outage Probability Analysis}
From the cascaded channel model, the instantaneous received SNR at time $t$ is
\begin{equation} \label{eq:inssnr}
  \gamma(t) = \gamma_0
  \left|\sum_{n \in N_{\mathrm{eff}}(t)} \eta_n(t)
  \exp\!\left(j\!\left(\theta_n(t)
  - \frac{2\pi}{\lambda}\rho_n(t)\right)\right)\right|^2,
\end{equation}
where $\gamma_0 = P_t / \sigma_n^2$ is the baseline SNR determined by transmit power and noise variance, $\theta_n(t)$ are RIS phase shifts, and $\rho_n(t)$ are path length differences depending on antenna positions $\mathbf{X}_n(t)$. With optimal phase alignment $\theta_n(t) = \frac{2\pi}{\lambda}\rho_n(t)$, the SNR simplifies to $\gamma(t) = \gamma_0 N_{\mathrm{eff}}(t)^2$ in the LoS-dominated regime. 

The outage probability is then computed by analyzing the statistical distribution using Moment-Generating Functions (MGFs), Jensen Inequality and Gaussian approximations. Let $Y(t) = N_{\text{eff}}(t)$, where $Y(t)$ is a random variable representing the effective number of contributing antennas at time $t$. Since, $\gamma(t) = \gamma_0 + 10\log_{10}(\text{max}(1, Y(t)))$, the outage event $\gamma(t) \leq \gamma_{th}$ is equivalent to, $Y(t) \leq \gamma_{\text{thresh}}$, where $\gamma_{\text{thresh}} = 10^{(\gamma_{th} - \gamma_0)/10}$, where $\gamma(t)$ is the instantaneous SNR, $\gamma_0$ is the baseline SNR offset. Thus $P_{\text{out}} = \mathcal{P}(Y(t) \leq \gamma_{\text{thresh}})$ where $\mathcal{P}$ denotes probability. We consider the MGF of $Y(t)$, $\mathcal{M}_Y(s) = \mathrm{E}[\exp(sY(t))]$, where $s$ is the Laplace constant using Chernoff bound, $\mathcal{P}(Y(t) \leq \gamma_{\text{thresh}}) \leq \exp(s \gamma_{\text{thresh}}) \, \mathcal{M}_Y(-s)\ \ \forall s > 0$ which denotes the upper bound of $P_{\text{out}}$, where $\mathcal{M}_Y(-s) = \mathrm{E}[\exp(-sY(t))]$ is the MGF evaluated at $-s$.
To evaluate $P_{\mathrm{out}}$, we define the effective number of antenna as,
\begin{equation}\label{eq:neff}
    N_{\text{eff}}(t) = \sum_{n = 1}^N a_n(t) = \sum_{n = 1}^N \exp(-d_n(t)^2),
\end{equation}
where $a_n(t)$ is the spatial illumination weighting or effective gain contribution of antenna $n$ at time $t$, modeled as an exponentially decaying function of its squared distance from the illumination center, $d_n(t)$ is the distance from antenna $n$ to the center of illumination at time $t$. Under the approximation of weak spatial correlation between antenna elements in  (\ref{eq:spa_corr}), we can write $ \mathcal{M}_Y(s) = \prod_{n = 1}^N \mathrm{E}[\exp(sa_n(t))]$, 
where each term in $\mathrm{E}[\exp(sa_n(t))]$ can be approximated by expanding $a_n(t)$ around its mean $\bar{a} = \mathrm{E}[\exp(a_n(t))]$, assuming small variance,
\begin{align}\label{e27}
    \mathrm{E}[\exp(sa_n(t))] \approx \exp\big(s\bar{a} + s^2\text{Var}(a_n(t))/2\big).
\end{align}
On the other hand, $P_{\text{out}}$ can be lower bounded using Jensen's inequality as, $\mathrm{E}[\log_{10}(Y(t))] \leq \log_{10}(\mathrm{E}[Y(t)])$ and average SNR results in $\mathrm{E}[\gamma(t)] \leq \gamma_0 + 10\log_{10}(\mathrm{E}[Y(t)])$.

Since $Y(t) = \sum_{n=1}^{N} a_n(t)$ is a sum of $N$ antenna effectiveness terms, where each $a_n(t) = \exp(-d_n(t)^2)$ depends on the independent stochastic trajectory of antenna $n$, the Central Limit Theorem implies that $Y(t)$ is approximately Gaussian for moderate $N$. The independence follows from the fact that each antenna evolves under its own SDE with independent Brownian motion. While the $a_n(t)$ are bounded in $[0,1]$ and not strictly Gaussian individually, their sum converges rapidly to a normal distribution. Let us assume $Y(t)$ is approximately Gaussian with mean $\mu_Y$ and variance $\sigma_Y^2$. Consequently, we get $P_{\text{out}}(\gamma_{th}) \approx \mathcal{Q}\Big(\frac{\mu_Y - \gamma_{\text{thresh}}}{\sigma_Y}\Big)$, where $\mathcal{Q}(\cdot)$ is the standard $\mathcal{Q}$-function, $\mathcal{Q}(z) = 1/\sqrt{2\pi}\int_z^{\infty} \exp(-r^2/2)\mathrm{d}r$. Since the $N$ antenna elements evolve under independent, identically distributed SDE, the effectiveness terms $\{a_n(t)\}_{n=1}^{N}$ are i.i.d.\ random variables with common mean $\bar{a} = \mathrm{E}[a_n(t)]$ and variance $\mathrm{Var}(a_n(t))$. By linearity of expectation and the variance property for independent sums, $\mu_Y = \mathrm{E}[Y(t)] = \mathrm{E}\!\left[\sum_{n=1}^{N} a_n(t)\right] = N\,\bar{a}$ and $\sigma_Y^2 = \mathrm{Var}(Y(t)) = \sum_{n=1}^{N} \mathrm{Var}(a_n(t)) = N\,\mathrm{Var}(a_n(t))$, where the second equality in each line uses the independence of $\{a_n(t)\}$. We can arrive at the final closed-form approximation,
\begin{align}\label{e29}
    P_{\text{out}}(\gamma_{th}) \approx \mathcal{Q}\Bigg(\frac{N\bar{a} - \gamma_{\text{thresh}}}{\sqrt{N\text{Var}(a_n(t))}}\Bigg)
\end{align}
where $\bar{a}$ and $\text{Var}(a_n(t))$ are functions of the movement randomness and optimization ($\sigma$, $V_{\text{drift}}$).

\subsection{Overhead-aware robust Two-timescale joint movement--phase optimization}
\label{subsec:practical_robust_overhead}

\begin{figure}[t]
\centering
\includegraphics[width=\columnwidth]{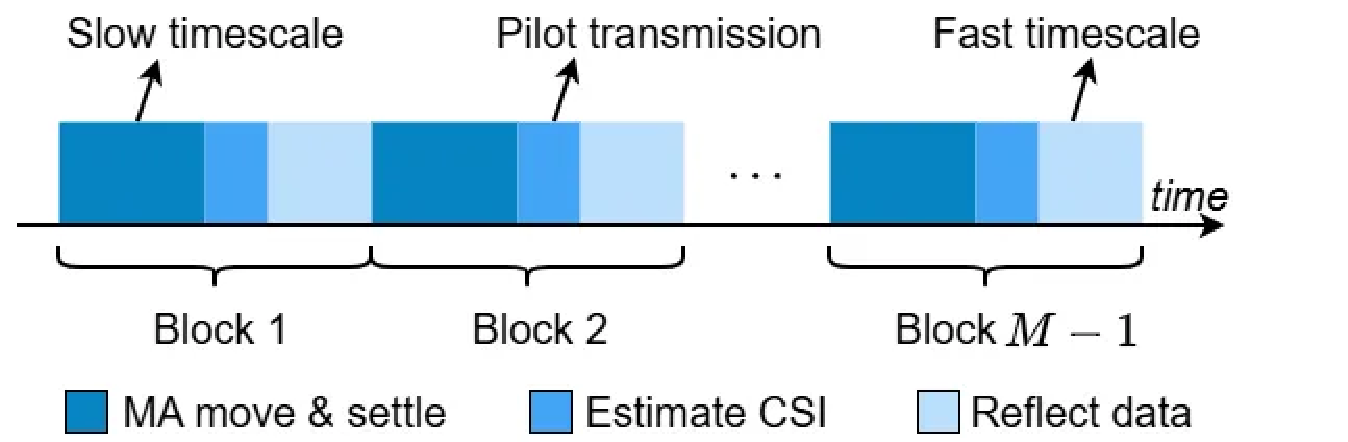}
\caption{Two-timescale move--estimate--reflect protocol illustration. Each coherence block is partitioned into three sequential phases: MA movement and settling (slow timescale), pilot-based CSI estimation, and data reflection (fast timescale). The protocol repeats across $M$ consecutive blocks.}
\label{fig:twoscale_model}
\end{figure}

\subsubsection{Two-timescale move--estimate--reflect protocol}
\label{subsubsec:two_timescale}

While ~(\ref{eq:inssnr}) models a continuous-time SNR $\gamma(t)$, practical operation requires separating mechanical movement from CSI acquisition and data reflection. We therefore partition time into coherence blocks indexed by $m\in\{0,1,\dots,M-1\}$.
Let $T_c^{(m)}$ denote the coherence time of block $m$ (e.g., $T_c^{(m)}\approx 1/(2\pi f_D^{(m)})$). Within block $m$, we use three sequential phases, illustrated in
Fig.~\ref{fig:twoscale_model}:
\begin{enumerate}
\item \textbf{Move \& settle (slow timescale):} during the the movement/settling time $\tau_{\rm mv}^{(m)}$, the MA elements move from their previous positions to target positions, then stop and settle. Let us denote the (now fixed) positions by $\{X_n^{(m)}\}_{n=1}^N$ at the end of this phase. 
\item \textbf{Estimate CSI:} during the pilot/estimation time $\tau_{\rm CSI}^{(m)}$, pilots are transmitted and the cascaded channel gains are estimated, yielding the estimated cascaded channel gains $\{\hat{\eta}_n^{(m)}\}_{n \in \mathcal{N}_{\mathrm{eff}}^{(m)}}$, where $\hat{\eta}_n^{(m)} \in \mathbb{C}$ is the pilot-based estimate of the cascaded complex gain $\eta_n^{(m)}$ for element $n$ in block $m$. $\mathcal{N}_{\mathrm{eff}}^{(m)} \triangleq
\{n \in \{1,\dots,N\} : a_n^{(m)} \geq a_{\min}\}$ is the set of effective elements, $N_{\mathrm{eff}}^{(m)} \triangleq
|\mathcal{N}_{\mathrm{eff}}^{(m)}|$ is its cardinality, and $a_{\min}$ is the minimum effectiveness threshold below which elements are negligible.
\item \textbf{Reflect data (fast timescale):} during the remaining time
$T_{\rm d}^{(m)}\triangleq T_c^{(m)}-\tau_{\rm mv}^{(m)}-\tau_{\rm CSI}^{(m)}$, antennas remain stationary
and the RIS applies fixed phases $\{\Theta_n^{(m)}\}$ to reflect data symbols.
\end{enumerate}
This protocol ensures that (i) channel estimation is performed under (approximately) static geometry, and (ii) coherent reflection uses phases optimized for the currently estimated channel.

The MA reconfiguration overhead in block $m$ consists of multiple physical components rather than a single lumped delay. Specifically, the movement time is decomposed as $\tau_{mv}^{(m)} = \tau_{tr}^{(m)} + \tau_{set}^{(m)} + \tau_{sync}^{(m)}$, where $\tau_{tr}^{(m)}$ is the trajectory execution time, $\tau_{set}^{(m)}$ is the mechanical settling time required before coherent reflection, and $\tau_{sync}^{(m)}$ is the controller/synchronization delay. With displacement $\Delta X^{(m)}_n = X^{(m)}_n - X^{(m - 1)}_n$, a practical model is $\tau_{tr}^{(m)} = \frac{\text{max}_n||\Delta X^{(m)}_n||}{v_{\text{max}}}$, where $v_{\max}$ is the maximum velocity of the MA. Likewise, the energy consumed by movement can be written as $E_{mv}^{(m)} = E_{act}^{(m)} + E_{set}^{(m)} + E_{ctrl}^{(m)}$, where $E_{act}^{(m)}$ is the actuation energy for physical displacement, $E_{set}^{(m)}$ is the stabilization energy, and $E_{ctrl}^{(m)}$ accounts for sensing and driver circuitry. Hence, longer MA displacements incur larger latency and energy costs, which motivates the overhead-aware trajectory design adopted in this work. 
\subsubsection{CSI overhead penalty (time available for data)}
\label{subsubsec:overhead}
The fraction of coherence time available for payload reflection in block $m$ is
\vspace{-2mm}
\begin{equation}
\kappa^{(m)} \triangleq \left[1-\frac{\tau_{\rm mv}^{(m)}+\tau_{\rm CSI}^{(m)}}{T_c^{(m)}}\right]^+,
\qquad [x]^+\triangleq\max\{x,0\},
\label{eq:kappa_block}
\end{equation}
where $T_c^{(m)}$ is the coherence time. Hence $\kappa^{(m)}\in[0,1]$ captures the
\emph{overhead penalty} due to movement and CSI acquisition. 

Since MA repositioning is performed before CSI acquisition and data reflection, the mechanical movement-and-settling interval $\tau^{(m)}_{mv}$ directly contributes to the per-block service latency. Together with the pilot overhead $\tau^{(m)}_{CSI}$, it reduces the remaining payload transmission duration to $T^{(m)}_{d} = T^{(m)}_{c} - \tau^{(m)}_{mv} - \tau^{(m)}_{CSI}$. Hence, from a URLLC perspective, antenna motion should be interpreted as a reliability-enhancing but latency-consuming operation: aggressive repositioning can improve SNR and outage performance, but excessive movement may violate tight delay budgets if $\tau^{(m)}_{mv}$ becomes comparable to the coherence time. Accordingly, the proposed overhead-aware controller favors movement only when the expected reliability gain justifies the associated latency cost; otherwise, it selects smaller adjustments or keeps the MA configuration unchanged. In this sense, the proposed scheme is compatible with URLLC only in the regime where the mechanical reconfiguration time is sufficiently smaller than the coherence time and the allowable end-to-end latency budget (packet deadline), i.e., $\tau^{(m)}_{mv} + \tau^{(m)}_{CSI} \ll T^{(m)}_{c}$ and below the allowable end-to-end latency budget.

A generic pilot model is
\begin{equation}
\tau_{\rm CSI}^{(m)} = N_p^{(m)} T_s,
\qquad
N_p^{(m)}=\xi\,K\,N_{\rm eff}^{(m)},
\label{eq:tau_csi_block}
\end{equation}
where $T_s$ is the symbol duration, $K$ is the number of served users (or links) to be estimated, and $\xi \geq 1$ captures protocol-dependent pilot overhead scaling. For orthogonal training where each effective element requires a dedicated pilot, $\xi = 1$; for grouped estimation schemes that estimate elements in subgroups of size $G$, $\xi = \lceil N_{\mathrm{eff}}^{(m)} / G \rceil / N_{\mathrm{eff}}^{(m)} \leq 1$ (reducing overhead); for feedback-assisted protocols requiring both uplink and downlink pilots, $\xi = 2$.

\subsubsection{Imperfect CSI model and robust SNR guarantee}
\label{subsubsec:robust}
In ~(\ref{eq:inssnr}), $\eta_n(t)\in\mathbb{C}$ denotes the cascaded (Tx$\to$MA-RIS$\to$Rx) complex gain for element $n$.
Under pilot-based estimation, we only observe an estimate:
\begin{equation}
\hat{\eta}_n^{(m)} = \eta_n^{(m)} + e_n^{(m)}, \qquad \forall n\in\mathcal{N}_{\rm eff}^{(m)},
\label{eq:eta_hat}
\end{equation}
where $e_n^{(m)}\in\mathbb{C}$ is the estimation error, and $\mathbf{e}^{(m)}\triangleq [e_n^{(m)}]_{n\in\mathcal{N}_{\rm eff}^{(m)}}$ is the aggregated error vector satisfying the norm-bounded uncertainty constraint $\|\mathbf{e}^{(m)}\|_2 \le \varepsilon^{(m)}$, with $\varepsilon^{(m)} \geq 0$ denoting the CSI error radius.
Under standard least-squares estimation with $N_p^{(m)}$ pilot symbols at pilot SNR $\gamma_p$, the per-element estimation error satisfies $\mathrm{E}[\|e_n^{(m)}\|^2] \leq \sigma_e^2
= 1/(N_p^{(m)} \gamma_p)$, yielding the norm bound
$\varepsilon^{(m)} = \sigma_e \sqrt{N_{\mathrm{eff}}^{(m)}}
= \sqrt{N_{\mathrm{eff}}^{(m)} / (N_p^{(m)} \gamma_p)}$. Thus $\varepsilon^{(m)}$ decreases with longer pilot sequences or higher pilot SNR, and increases with more effective antennas requiring estimation.

Let us define the controllable phase term in block $m$ for element $n$ as $\phi_n^{(m)} \triangleq \Theta_n^{(m)} - \frac{2\pi}{\lambda}\rho_n^{(m)}$, where $\Theta_n^{(m)}\in[0,2\pi)$ is the RIS phase shift, $\lambda$ is the carrier wavelength, and $\rho_n^{(m)}$ is the path-length difference between the Tx and the reference~point~$\mathbf{r}_0$, induced by the fixed position $X_n^{(m)}$ of element $n$ in block $m$.
Using estimated CSI, define the coherent combining sum 
\begin{equation}
S^{(m)} \triangleq \sum_{n\in\mathcal{N}_{\rm eff}^{(m)}} \hat{\eta}_n^{(m)} e^{j\phi_n^{(m)}} .
\label{eq:S_block}
\end{equation}
The true (unknown) sum equals $\sum_{n} \eta_n^{(m)} e^{j\phi_n^{(m)}} = S^{(m)} - \sum_n e_n^{(m)} e^{j\phi_n^{(m)}}$. Let $\mathbf{w}^{(m)}\triangleq [e^{j\phi_n^{(m)}}]_{n\in\mathcal{N}_{\rm eff}^{(m)}}$ so that $\|\mathbf{w}^{(m)}\|_2=\sqrt{N_{\rm eff}^{(m)}}$ (since $|e^{j\phi}|=1$). Then, by the reverse triangle inequality and Cauchy--Schwarz,
\begin{align}
\left|\sum_{n\in\mathcal{N}_{\rm eff}^{(m)}} \eta_n^{(m)} e^{j\phi_n^{(m)}}\right|
&\ge |S^{(m)}| - \|\mathbf{e}^{(m)}\|_2\,\|\mathbf{w}^{(m)}\|_2
\nonumber\\
&\ge |S^{(m)}| - \varepsilon^{(m)}\sqrt{N_{\rm eff}^{(m)}}.
\label{eq:robust_bound}
\end{align}
Substituting \eqref{eq:robust_bound} into the SNR structure of ~(\ref{eq:inssnr})
yields a conservative \emph{robust SNR}:
\begin{equation}
\gamma_{\rm rob}^{(m)}
\triangleq
\gamma_0\left(\left[\,|S^{(m)}|-\varepsilon^{(m)}\sqrt{N_{\rm eff}^{(m)}}\,\right]^+\right)^2,
\label{eq:gamma_rob_block}
\end{equation}
~\eqref{eq:gamma_rob_block} shows that imperfect CSI introduces a robustness margin
proportional to the uncertainty radius $\varepsilon^{(m)}$ and $\sqrt{N_{\rm eff}^{(m)}}$.

Combining the overhead penalty in \eqref{eq:kappa_block} with the robust SNR in
\eqref{eq:gamma_rob_block}, we define the \emph{net robust} SNR in block $m$ as:
\begin{equation}
\gamma_{\rm net,rob}^{(m)} \triangleq \kappa^{(m)}\,\gamma_{\rm rob}^{(m)}.
\label{eq:gamma_net_rob}
\end{equation}
The joint movement--phase design then aims to maximize the time-average expected net robust SNR:
\begin{equation}
\max_{\{{V}_n(t),\,\Theta_n^{(m)}\}}
\ \mathrm{E}\!\left[\frac{1}{M}\sum_{m=0}^{M-1}\gamma_{\rm net,rob}^{(m)}\right],
\label{eq:objective_unified}
\end{equation}
subject to the same mobility dynamics and feasibility constraints as in  (\ref{eq:sde}),
with the additional protocol constraint that antennas are stationary during CSI acquisition and data reflection within each coherence block (movement is confined to the \emph{move \& settle} phase).

The objective function $\gamma_{\rm net,rob}^{(m)}$ is non-convex and the movement control and phase design are interdependent. Thus the problem is a non-convex, stochastic control problem. We next use HJB formulation as the foundational tool in continuous-time stochastic optimal control.

Let the system state be the concatenated vector of all MA positions at time $t$, $Z(t) = [Z_1(t), Z_2(t), \dotso, Z_N(t)]^T \in \mathbb{R}^{2N}$. We also define the value function $\mathcal{V}(Z, t)$ as the maximum expected cumulative reward from time $t$ onwards that can be expressed as:
\begin{align}\label{e33}
    \mathcal{V}(Z, t) = \max_{V(\cdot), \theta(\cdot)} \mathrm{E} \bigg[\int_t^{\tau} \gamma(s)\mathrm{d}s|Z(t) = Z\bigg]
\end{align}
where $V(t) = \{{V}_n(t)\}_{n = 1}^N$, $\theta(t) = \{\theta_n(t)\}_{n = 1}^N$ and $\gamma(t)$ is the instantaneous SNR expressed using (\ref{eq:inssnr}). In the discrete dynamic programming formulation, the stage reward at time step $k$ (interpreted as a block index) is updated from $r_k=\gamma_k$ to the overhead-aware robust reward $\gamma_{\mathrm{net,rob},k}$:
\vspace{-1mm}
\begin{equation}
r_k(Z_k,\theta_k) = \gamma_{{\rm net,rob},k}
= \kappa_k\,\gamma_0\left(\left[\,|S_k|-\varepsilon_k\sqrt{N_{\rm eff}^k}\,\right]^+\right)^2,
\label{eq:dp_reward_unified}
\end{equation}
where
$S_k \triangleq \sum_{n=1}^{N_{\rm eff}^k}\hat{\eta}_n^k
\exp\!\left(j(\theta_n^k-\frac{2\pi}{\lambda}\rho_n^k)\right)$,
$\kappa_k \triangleq \left[1-\frac{\tau_{{\rm mv},k}+\tau_{{\rm CSI},k}}{T_{c,k}}\right]^+$,
and $\varepsilon_k$ is the CSI uncertainty radius at step $k$. The HJB equation governing $\mathcal{V}(Z, t)$, is expressed as,
\begin{align}\label{e34}
    \frac{\delta V}{\delta t} + \max_{V(t), \theta(t)}\Bigg[\nabla_Z\mathcal{V} + \frac{\sigma^2}{2}\tr\Big(\nabla_Z^2\mathcal{V}\Big) + \gamma(Z, \theta)\Bigg] = 0
\end{align}
where $\nabla_Z$ denotes the control strategy with terminal condition $\mathcal{V}(Z,T) = 0$. Solving ~(\ref{e34}) provides the optimal control law $(V^*, \theta^*)$ as a function of the current state of the MA elements. Now, in this case, the solution state-space dimension is $2N$, making ~(\ref{e34}) harder to solve as $N$ increases. With even moderate $N = 10$, the joint state $Z(t) = [Z_1(t), \dots, Z_N(t)]^T \in \mathbb{R}^{2N}$
has dimension $2N = 20$. Including time as an additional variable, the value function $V(Z,T)$ is defined over $\mathbb{R}^{2N+1} = \mathbb{R}^{21}$, which is  intractable for grid-based solvers. Therefore, we now resort to Discrete Dynamic Programming which can provide a grid-based approximation by applying backward recursion.

The continuous-time HJB formulation in~(\ref{e34}) is defined on the joint state
$Z(t)=[Z_1(t),\ldots,Z_N(t)]^T\in\mathbb{R}^{2N}$, where $Z_n(t)\in\mathbb{R}^2$ denotes the 2D position of antenna $n$.
A direct grid-based Dynamic Programming (DP) discretization of this joint state would require a grid of size $G^N$ if each antenna is discretized over $G$ candidate positions, and a joint action space of size $A^N$ if each antenna has $A$ candidate velocity actions. 
Hence, the complexity per stage scales as $\mathcal{O}(G^N A^N)$, which is computationally prohibitive for moderate-to-large $N$. Therefore, the discrete method used in our simulations is \emph{not} a centralized $2N$-dimensional DP.

\subsubsection{Decomposition used: closed-form phase update + sequential per-antenna DP}
\label{subsubsec:decomposition}

Under the move--estimate--reflect protocol (Section~III-C), during each block the $k$ antennas are stationary during CSI estimation and data reflection, so the geometry-dependent terms $\{\rho_n^k\}$ are fixed within the block.
Given the estimated cascaded gains $\{\hat{\eta}_n^k\}$, the phase shifts that maximize coherent combining are obtained by aligning the reflected terms:
\begin{equation}
\theta_{n}^{k\star} \;=\; \frac{2\pi}{\lambda}\rho_n^k \;-\; \arg(\hat{\eta}_n^k) \;+\; \psi_k,
\qquad n=1,\ldots,N_{\mathrm{eff}}^k,
\label{eq:phase_closed_form_revised}
\end{equation}
where $\psi_k\in[0,2\pi)$ is an arbitrary common offset (it does not change the magnitude of the coherent sum).
This eliminates the need for an $N$-dimensional grid search over $\theta^k$. The discrete phase alphabet $\mathcal{M} = \{0, 2\pi/M, \dots, 2\pi(M-1)/M\}$ introduced for quantized phase shifts is therefore not
required in the implemented solver, as the closed-form solution (\ref{eq:phase_closed_form_revised}) provides the optimal continuous-valued phases directly.

To handle the remaining high-dimensional movement optimization, we adopt a block-coordinate update over antennas.
Let $z_{n,k}\in\mathbb{R}^2$ denote the position of antenna $n$ at block $k$, and let $\mathcal{G}\subset\mathbb{R}^2$
be a 2D grid of candidate positions with $|\mathcal{G}|=G$. Crucially,  this grid discretizes each individual antenna position $z_{n,k}$, not the full joint state $Z_k\in\mathbb{R}^{2N}$. At outer iteration $i$, antennas are updated sequentially while holding all others fixed at their latest trajectories. When updating antenna $n$, we solve a 2D DP over $\mathcal{G}$ with dynamics:
\begin{equation}
z_{n,k+1} = z_{n,k} + v_{n,k}\Delta t + \sigma\sqrt{\Delta t}\,\varepsilon_{n,k},
\label{eq:per_antenna_transition_revised}
\end{equation}
where $z_{n,k+1} \in \mathbb{R}^2$ is the updated position at the next block $k+1$, $v_{n,k} \in \nu$ is the velocity control input selected from a finite action set $\nu$, $\Delta t$ is the duration of one coherence block, $\sigma$ is the diffusion coefficient defined in~(\ref{eq:sde}), and $\varepsilon_{n,k} \sim \mathcal{N}(\mathbf{0}, \mathbf{I}_2)$ is a standard 2D Gaussian random vector representing the discretized Brownian motion increment.

The overhead-aware robust stage reward $\gamma_{\mathrm{net,rob},k}$, depends on the coherent sum $S_k$ and the overhead factor $\kappa_k$. Under the closed-form phase alignment in~\eqref{eq:phase_closed_form_revised}, the coherent sum magnitude is given by $|S_k| \;=\; \sum_{m=1}^{N_{\mathrm{eff}}^k} |\hat{\eta}_m^k|$, so the reward depends on the \emph{sum of per-antenna magnitudes}, where each $|\hat{\eta}_m^k|$ is determined by geometry (and hence by the antenna position). During the update of antenna $n$ at iteration $i$, the fixed contribution of all other antennas is defined as $A^{(i)}_{k,-n}\triangleq \sum_{m\neq n} |\hat{\eta}_m^k(z_{m,k}^{(i)})|$.
Then the per-antenna DP uses the following coordinate surrogate reward:
\begin{equation}
\tilde{r}^{(i)}_{k,n}(z_{n,k},v_{n,k})
\triangleq
\kappa_k\,\gamma_0 \left( \left[ \Gamma^{(i)}_{k,n}(z_{n,k}) \right]^+ \right)^2,
\label{eq:coordinate_reward}
\end{equation}
where $\Gamma^{(i)}_{k,n}(z) = A^{(i)}_{k,-n} + |\hat{\eta}_n^k(z)| - \varepsilon_k\sqrt{N_{\mathrm{eff}}^k}$ is the robust array gain at iteration $i$ when antenna $n$ is at position $z$. This coordinate surrogate is a guaranteed lower bound consistent with the robust reward in ~(\ref{eq:dp_reward_unified}), while enabling tractable 2D DP updates.

The per-antenna \emph{cost-to-go} (or value function), which represents the maximum expected cumulative reward achievable from grid point $z$ at block $k$ onwards, is updated via the Bellman recursion:
\begin{align}
J^{(i)}_{k,n}(z) &= \max_{v\in\nu} \Big\{ \tilde{r}^{(i)}_{k,n}(z,v) + \mathrm{E}_{\varepsilon}\big[J^{(i)}_{k+1,n}(z')\big] \Big\}, \label{eq:per_antenna_dp_revised} 
\end{align}
with $z'$ given by ~\eqref{eq:per_antenna_transition_revised}, $J^{(i)}_{k+1,n}(z')$ denotes the expected future value from the successor state $z'$ at block $k+1$, and $J^{(i)}_{K,n}(z)=0 , \forall z$ is the terminal condition (no future reward beyond the final block $K$). After updating all $N$ antennas sequentially via coordinate descent (one full iteration $i$), phases are updated by ~\eqref{eq:phase_closed_form_revised}. This yields a fully reproducible algorithm that scales linearly in $N$ (up to a small number of outer iterations).

\subsubsection{Complexity analysis}
\label{subsubsec:complexity_revised}
Let $G=|\mathcal{G}|$ be the number of 2D grid points per antenna (e.g., $G=100$ for a $10\times10$ grid), $A=|\nu|$ the number of candidate velocity actions per antenna, $K$ the time horizon in blocks, $S$ the number of Monte Carlo samples used to approximate $\mathrm{E}_{\varepsilon}[\cdot]$, and $I$ the number of outer sequential-update iterations, in which one full pass updates all $N$ antennas sequentially (antenna 1, then 2, \dots, up to $N$), with each antenna's policy optimized while holding all other antennas' policies fixed at their latest values. A joint grid-based DP would require $\mathcal{O}(K\,G^N\,A^N)$ operations per sweep (curse of dimensionality), which is infeasible. Each per-antenna DP sweep costs $\mathcal{O}(K\,G\,A\,S)$, and we perform $N$ antenna updates over $I$ outer iterations. Thus, the overall complexity is $\mathcal{O}\!\left(I\,N\,K\,G\,A\,S\right)$, with memory requirement $\mathcal{O}(G)$ per antenna for storing the cost-to-go values $J^{(i)}_{k,n}$ at each grid point (or $\mathcal{O}(K G)$ if storing value functions across all time stages). For the simulation setting ($N=16$ and $G=100$), this is practical, whereas a centralized DP over $Z_k\in\mathbb{R}^{2N}$ would be computationally prohibitive.

\subsubsection{Discretized Space-Time Model}
We discretize the continuous dynamics  over an observation duration $T$ into a grid-based space-time model with:
\begin{align}
    \text{time}~&:~t_k = k\Delta t,\ \ k = 0, 1, \dotso, K,\nonumber\\
    &\text{where}~T = K\Delta t~\text{and}~\Delta t~\text{is the increment in time} \nonumber\\
    \text{state}~&:~\text{MA positions}~Z^k_n \in \mathcal{G} \subset \mathbb{R}^2 \nonumber\\
    &\text{sampled over a finite grid}, \nonumber\\
    \text{drift}~&:~V_n^k \in \nu~\text{(a finite set of velocity vectors) and}\nonumber\\
    \text{phase-shift}~&:~\theta_n^k \in \mathcal{M} \subset [0, 2\pi),~\text{discretized into $M$ levels.} \nonumber 
\end{align}
The state transition of each MA is then given by,
\begin{align}
    Z_n^{k+1} = Z_n^k + V_n^k \Delta t + \sigma \sqrt{\Delta t} \varepsilon^k_n
    \label{eq:statetran}
\end{align}
where $\varepsilon^k_n \in \mathcal{N}(0, \mathbf{I}_2)$ is the random noise due to mobility sampled during expectation calculations. In order to evaluate the resultant optimization, we generate results through simulation over a discrete panel of $10 \times 10$ grid and movement over $k = 10$ time steps. The stage reward at time step $k$ is given by the overhead-aware robust SNR in~(\ref{eq:dp_reward_unified}), i.e., $r_k(Z_k,\theta_k)=\gamma_{\mathrm{net,rob},k}$, where $Z^k = [\mathbf{X}_1^k, \dots, \mathbf{X}_N^k]$ is the concatenated position vector of all $N$ antennas at step $k$, and $\theta^k = [\theta_1^k, \dots,
\theta_N^k]^T$ are the corresponding RIS phase shifts.

Having defined the stage reward $r_k$ and the state transition (\ref{eq:statetran}), the remaining task is to determine the $(V^*, \theta^*)$ control law that maximizes the total expected reward over the horizon $k = 0, \dots, K$. In dynamic programming, the \emph{cost-to-go function} (also called the value function) $\mathcal{J}_k(Z^k)$ represents the maximum expected cumulative reward achievable from system state $Z^k$ at time step $k$ through the end of the horizon. It can be recursively defined via the Bellman equation as,
\begin{align}
    \mathcal{J}_k(Z^k) = \max_{V^k, \theta^k}\Big[r_k(Z^k, \theta^k) + \mathrm{E}_{\varepsilon}\Big[\mathcal{J}_{k+1}\big(Z^{k+1}\big)\Big]\Big]
\end{align}
with terminal condition $\mathcal{J}_k(Z^k) = 0$. Here $\max_{V^k, \theta^k}$ indicates that the expression is maximized over the control inputs at time step $k$, $V^k$ is the set of velocity vectors (drift) for the MA elements at time step $k$, $\theta^k$ is the phase shifts applied by the RIS elements at time step $k$, where $r_k(Z^k, \theta^k)$ is the stage reward at time step $k$, defined as the overhead-aware robust SNR in~(\ref{eq:dp_reward_unified}). $\mathrm{E}_{\varepsilon}[\cdot]$ denotes the expectation taken with respect to the random noise $\varepsilon_n^k$ due to mobility, which is sampled during expectation calculations, $\mathcal{J}_{k+1}(Z^{k+1})$ is the cost-to-go function for the next time step, $k+1$, with the state $Z^{k+1}$. The state $Z^{k+1}$ evolves based on the current state, velocity, and random noise.

\section{Performance Analysis}

The simulation setup reflects a sub-6~GHz MA-RIS deployment at carrier frequency $f = 3$~GHz ($\lambda = 0.1$~m). Table~\ref{tab:params} summarizes the simulation parameters. The antenna dynamics are simulated using the SDE with a normalized time step $\Delta t = 0.5$ s over a duration of $T_{\mathrm{d}} = 50$ s. $\tau_{\mathrm{mv}}$ is fixed at 0.15~s per block, representing a conservative time budget allocated for MA reconfiguration regardless of actual displacement. This ensures predictable latency compatible with URLLC constraints. Beyond the baseline configuration, the diffusion coefficient $\sigma$ is varied from $0.2-1.5$ to capture different levels of stochastic perturbation in the antenna motion, ranging from weak to diffusion-dominated dynamics. The drift velocity $V_{\mathrm{drift}}$ is varied from 0.5 to 2.5 m/s to evaluate the tradeoff between deterministic motion and stochastic diffusion. These parameter ranges are chosen to explore representative regimes of the antenna dynamics and to assess the robustness of the control strategies under different levels of motion uncertainty.
\vspace{-3mm}
\begin{table}[h]
\centering
\caption{Simulation Parameters}
\label{tab:params}
\begin{tabular}{|>{\raggedright\arraybackslash}p{0.35\linewidth}|l |l|}
\hline
\textbf{Parameter} & \textbf{Symbol} & \textbf{Value} \\
\hline
Carrier frequency        & $f$                          & 3 GHz ($\lambda = 0.1$ m)\\\hline
Number of antennas       & $N$                          & 16\\\hline
Baseline SNR             & $\gamma_0$                   & 15 dB\\\hline
Total duration            & $T_{\mathrm{d}}$& 50 s\\\hline
 Coherence time& $T_c$&1 s\\\hline
Time step                & $\Delta t$                   & 0.5 s\\\hline
Panel radius             & $R$                          & 2 m\\
\hline
\end{tabular}
\vspace{-3mm}
\end{table}

\subsection{Mathematical validation}  
\subsubsection{Distance Evolution Analysis}
\begin{figure}[t]
    \centering
    \makebox[\linewidth][h]{%
        \includegraphics[width=1.0\linewidth]{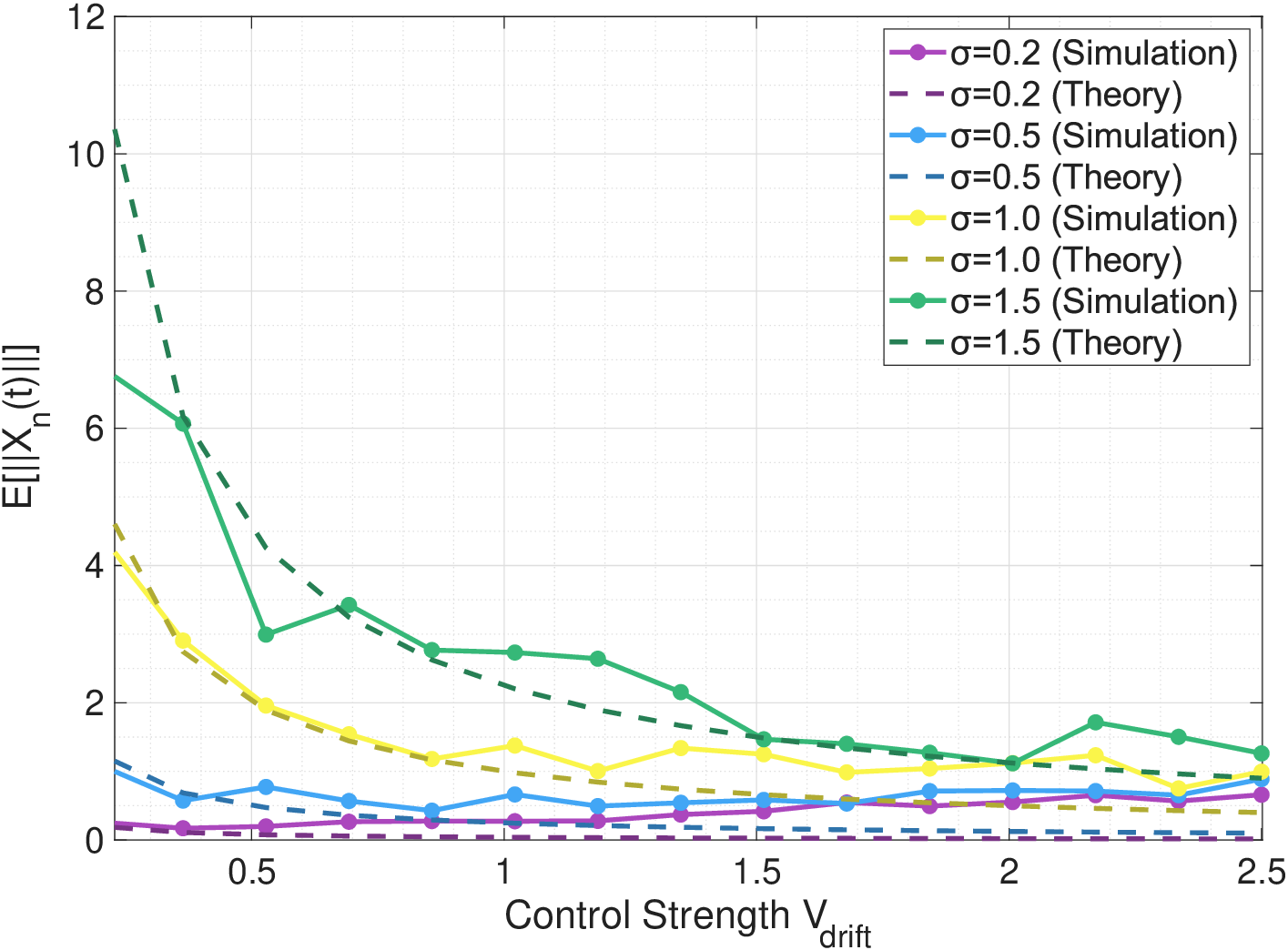}}
    \caption{Steady-state distance validation comparing theoretical prediction $\mathrm{E}[\|\mathbf{X}_n(t)\|] = \sigma^2/V_{\text{drift}}$ against Monte Carlo simulation for MA-RIS system with $N=16$ antennas. Four diffusion coefficients are tested ($\sigma \in \{0.2, 0.5, 1.0, 1.5\}$) across control strengths $V_{\text{drift}} \in [0.2, 2.5]$ m/s. Solid lines with markers represent empirical mean distance, while dashed lines show analytical predictions from  (\ref{e25}).}
    \label{math:vali1}
\end{figure}

Fig.~\ref{math:vali1} validates the theoretical model for expected distance evolution in Eq.~(\ref{e25}). The plot compares simulation results against the theoretical prediction $\mathrm{E}[\|\mathbf{X}_n(t)\|] = \sigma^2/V_{\text{drift}}$ for diffusion coefficients ($\sigma = 0.2, 0.5, 1.0, 1.5$) across control strengths $V_{\text{drift}} \in [0.2, 2.5]$, with a zoomed view up to $V_{\text{drift}}=2.4$ for clarity. The results demonstrate excellent agreement between simulation and theory, particularly for $V_{\text{drift}} > 0.5$. Two fundamental trends emerge:
\textbf{Impact of Diffusion:} When $V_{\text{drift}}$ remains constant, increasing $\sigma$ leads to larger expected distances from the illumination center. At $V_{\text{drift}} = 0.5$, expected distance increases from ~0.08 units ($\sigma = 0.2$) to 4.5 units ($\sigma = 1.5$). This quadratic dependence ($\mathrm{E}[\|\mathbf{X}_n\|] \propto \sigma^2$) indicates positioning accuracy degrades rapidly as environmental disturbances intensify.
\textbf{Impact of Control Strength:} When $\sigma$ is fixed, increasing $V_{\text{drift}}$ reduces expected distance. At $\sigma = 1.0$, distance decreases from ~5.0 units ($V_{\text{drift}} = 0.2$) to 0.4 units ($V_{\text{drift}} = 2.5$). However, beyond $V_{\text{drift}} \approx 1.5$, curves flatten, showing diminishing returns as the system approaches drift-diffusion equilibrium. This validates our theoretical framework: stronger drift control pulls antennas closer while increased randomness pushes them away, with $V_{\text{drift}} \sim \mathcal{O}(\sigma)$ representing an efficient operating point balancing control effort against positioning performance.
\begin{figure}
    \centering
    \makebox[\linewidth][c]{%
        \includegraphics[width=1.03\linewidth]{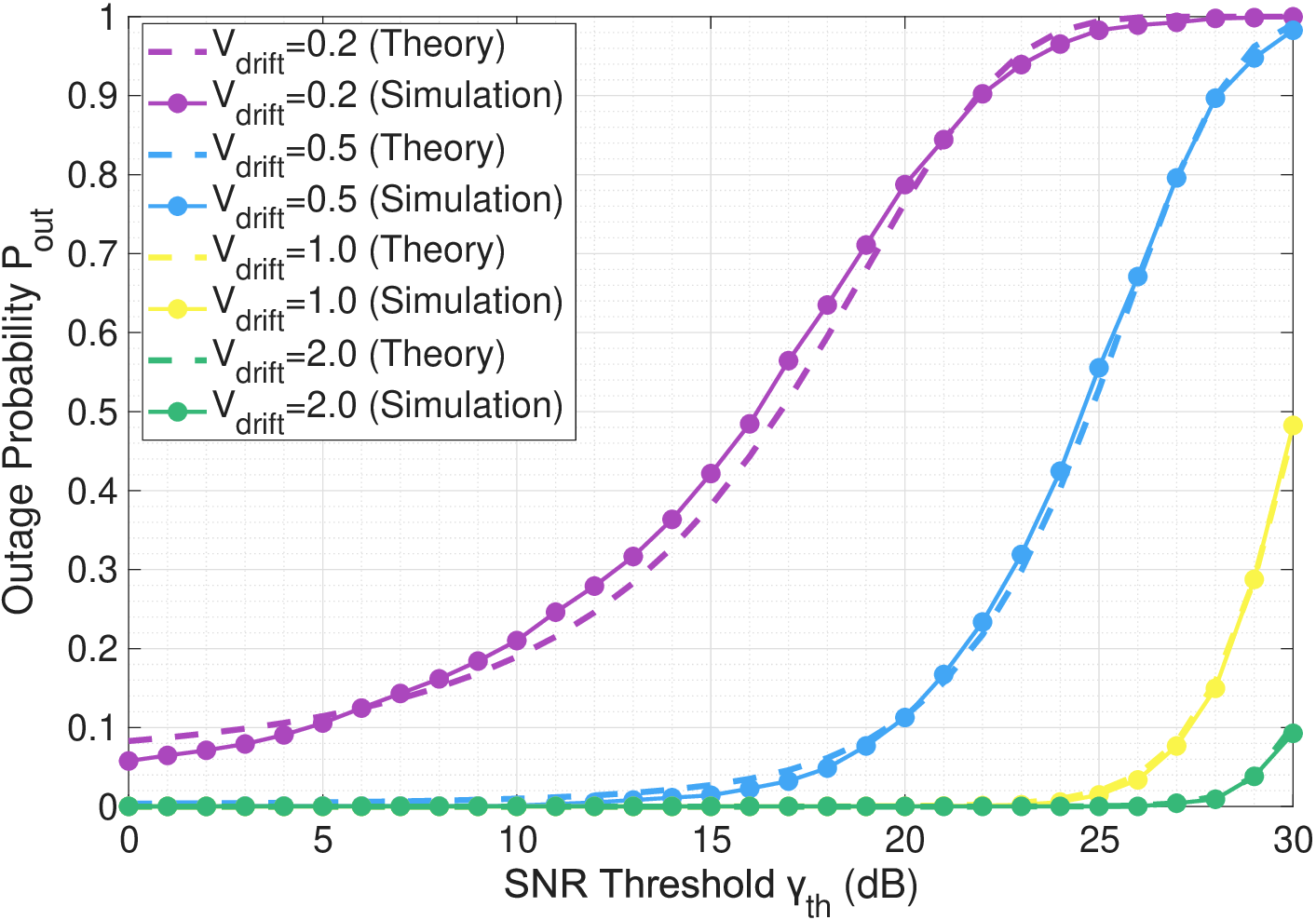}}
    \caption{Outage probability validation comparing Gaussian approximation (\ref{e29}) with Monte Carlo simulation for varying SNR thresholds $\gamma_{\text{th}} \in [0, 30]$ dB with diffusion coefficient $\sigma = 1.0$. Four control policies are evaluated: $V_{\text{drift}} \in \{0.2, 0.5, 1.0, 2.0\}$. Dashed lines represent theoretical curve using (\ref{e29}), while solid lines with markers show empirical outage frequencies.}
        \label{math:vali2}
        \vspace{-3mm}
\end{figure}
\subsubsection{Outage Probability Analysis}
\begin{figure}[t]
    \centering
    \makebox[\linewidth][c]{%
        \includegraphics[width=1\linewidth]{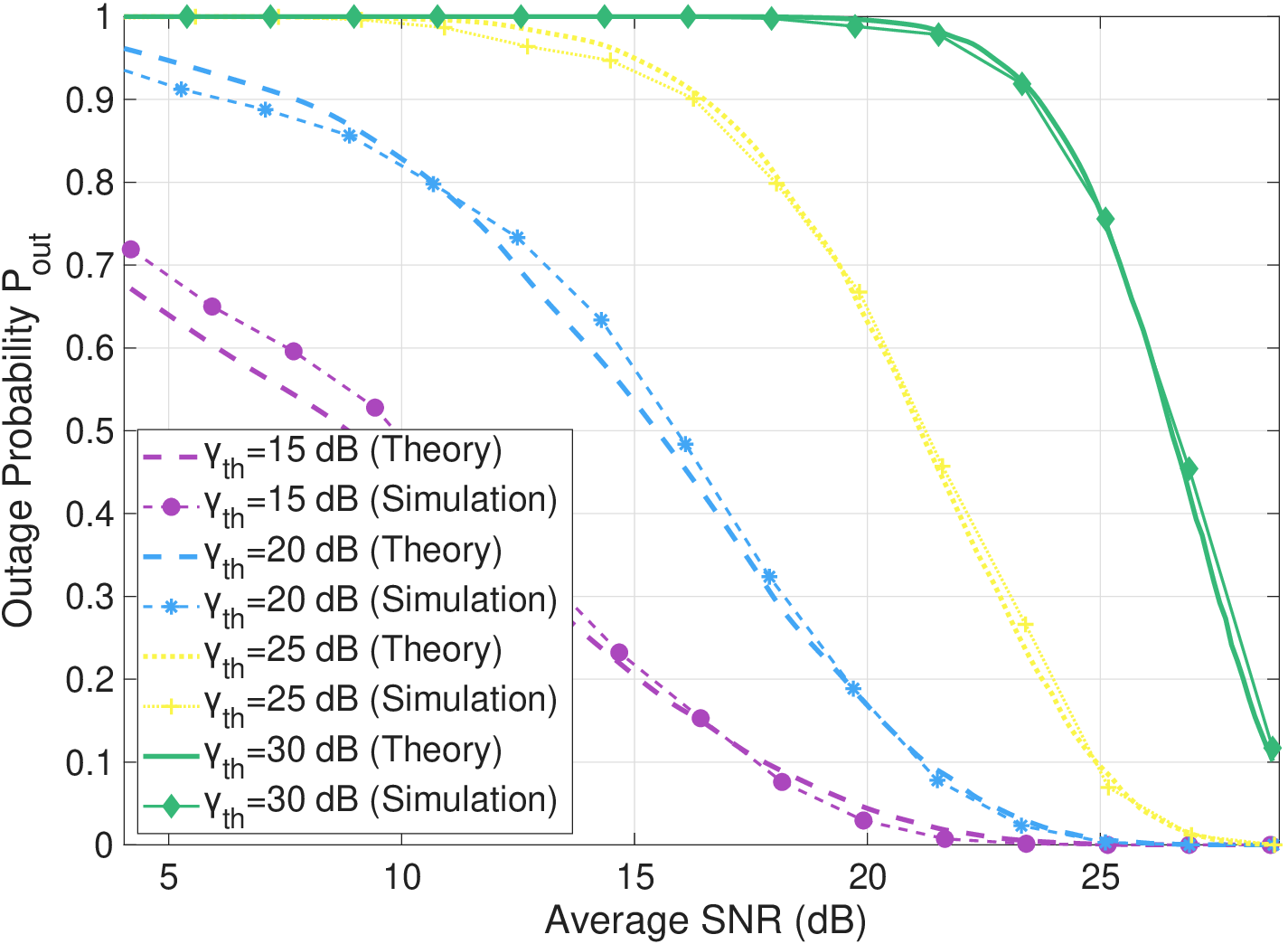}}
    \caption{Outage probability verus the average SNR for four fixed threshold levels: $\gamma_{th} \in \{15, 20, 25, 30\}$ dB with $\sigma = 1.0$ and $V_{drift} = 1.0$. Average SNR is swept across $[8, 32]$ dB by varying baseline SNR offset $\gamma_0$. Theoretical curves (dashed) computed using  (\ref{e29}) closely match simulation (solid).}
    \label{math:avgsnr}
    \vspace{-3mm}
\end{figure}

Figs.~\ref{math:vali2} and \ref{math:avgsnr} validate Eq.~(\ref{e29}) by examining outage probability versus SNR threshold and average SNR. Fig.~\ref{math:vali2} compares theoretical predictions against Monte Carlo simulations for various drift control strengths ($V_{\text{drift}} = 0.2, 0.5, 1.0, 1.5$). Lower control strengths ($V_{\text{drift}} = 0.2$) produce steeper outage curves, with $P_{\text{out}}$ rising from 0.1 to 0.9 within a narrow 5 dB window (10-15 dB threshold range), indicating high sensitivity when drift is weak. Stronger control ($V_{\text{drift}} = 2.0$) yields gentler slopes, maintaining $P_{\text{out}}$ near zero until $\gamma_{th}$ exceeds 25 dB. Fig.~\ref{math:avgsnr} further examines outage for SNR thresholds $\gamma_{th} = 15, 20, 25, 30$ dB. For $\gamma_{th} = 15$ dB, outage decreases from ~0.7 at 5 dB to below 0.05 at 23 dB. Higher thresholds shift transitions rightward, with $\gamma_{th} = 30$ dB maintaining outage above 0.1 even at 30 dB average SNR. The close match between theoretical Q-function approximation and simulation confirms our MGF-based approach accurately captures $N_{\text{eff}}(t)$ statistics. This demonstrates the reliability-SNR tradeoff: maintaining fixed outage under stricter thresholds requires proportionally higher baseline SNR.
\subsection{Parameter Sensitivity Analysis }
\subsubsection{Impact of Diffusion Coefficient $\sigma$}
\begin{figure}[t]
    \centering
    \makebox[\linewidth][c]{%
        \includegraphics[width=0.99\columnwidth]{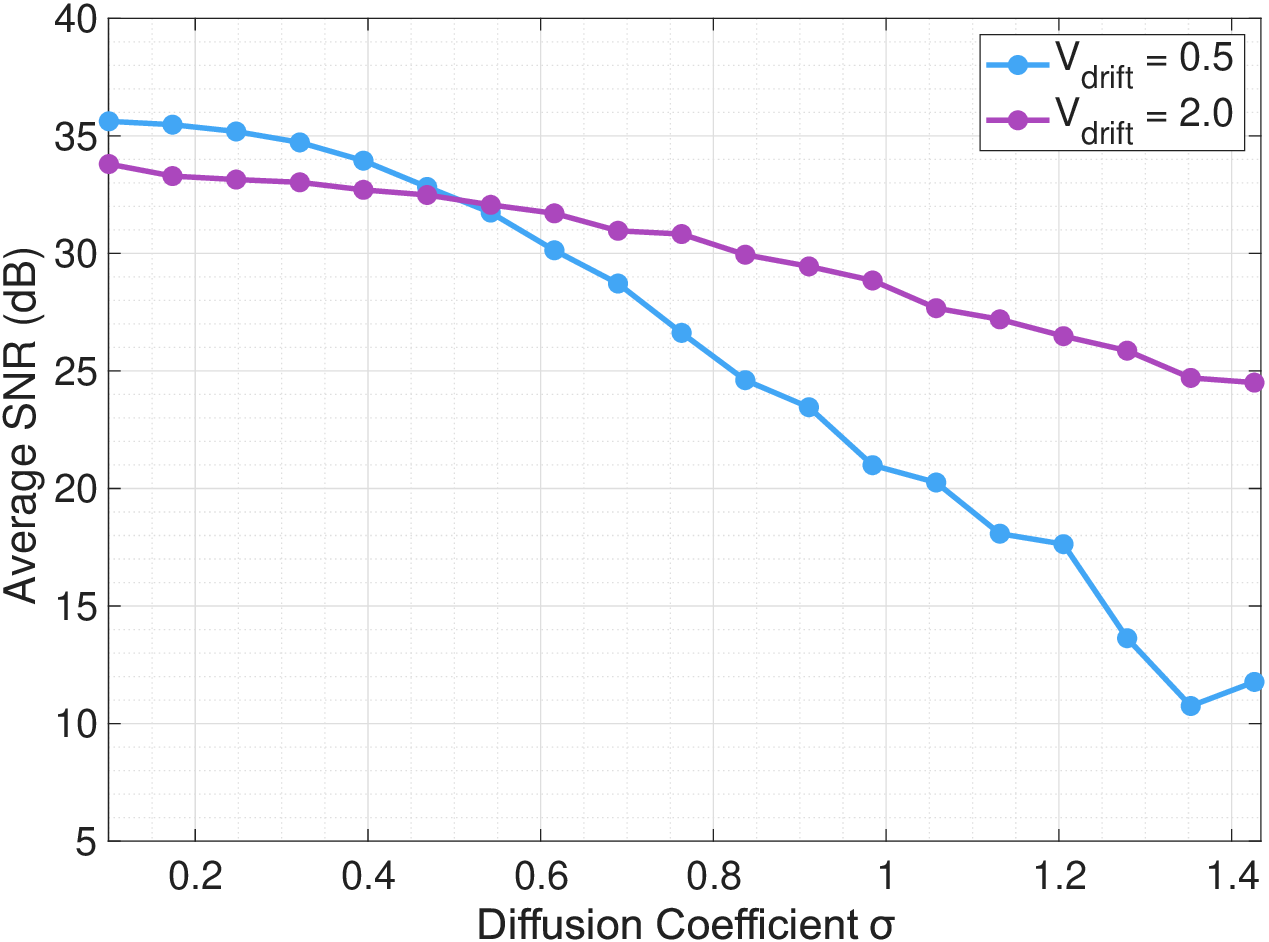}}
    \caption{Impact of environmental diffusion on average received SNR for  $V_{\text{drift}} \in \{0.5, 2.0\}$. Diffusion coefficient swept across $\sigma \in [0.1, 1.5]$ representing increasing mechanical noise and environmental perturbations.}
    \label{se:sigma}
\end{figure}

Fig. \ref{se:sigma} illustrates the sensitivity of average SNR to the diffusion coefficient $\sigma$ for two drift control settings ($V_{drift} = 0.5, 2.0$). The average SNR exhibits a monotonically decreasing relationship with $\sigma$, degrading from approximately 35 dB at $\sigma = 0.1$ to 23 dB at $\sigma = 1.5$ for $V_{drift} = 2.0$, and from 34 dB to 7 dB for $V_{drift} = 0.5$. This behavior directly reflects  (\ref{e25}): the steady-state distance $\mathrm{E}\big[||\mathbf{X}_n(t)||\big] = \sigma^2/V_{\text{drift}}$ increases quadratically with $\sigma$,  dispersing antennas from the illumination center. The stark contrast between $V_{drift} = 0.5$ and $V_{drift} = 2.0$ curves reveals that stronger control provides greater resilience against diffusion-induced performance degradation. With $V_{drift} = 2.0$, the system maintains SNR above 20 dB even at $\sigma= 1.5$, while $V_{drift} = 0.5$ suffers a very sharp degradation below 10 dB at the same diffusion level. This quantifies the critical importance of sufficient drift control strength in maintaining system performance under environmental perturbations or mechanical instabilities.

\subsubsection{Impact of Control Strength $V_{drift}$}
\begin{figure}[t]
    \centering
    \makebox[\linewidth][c]{%
            \includegraphics[width=0.99\linewidth]{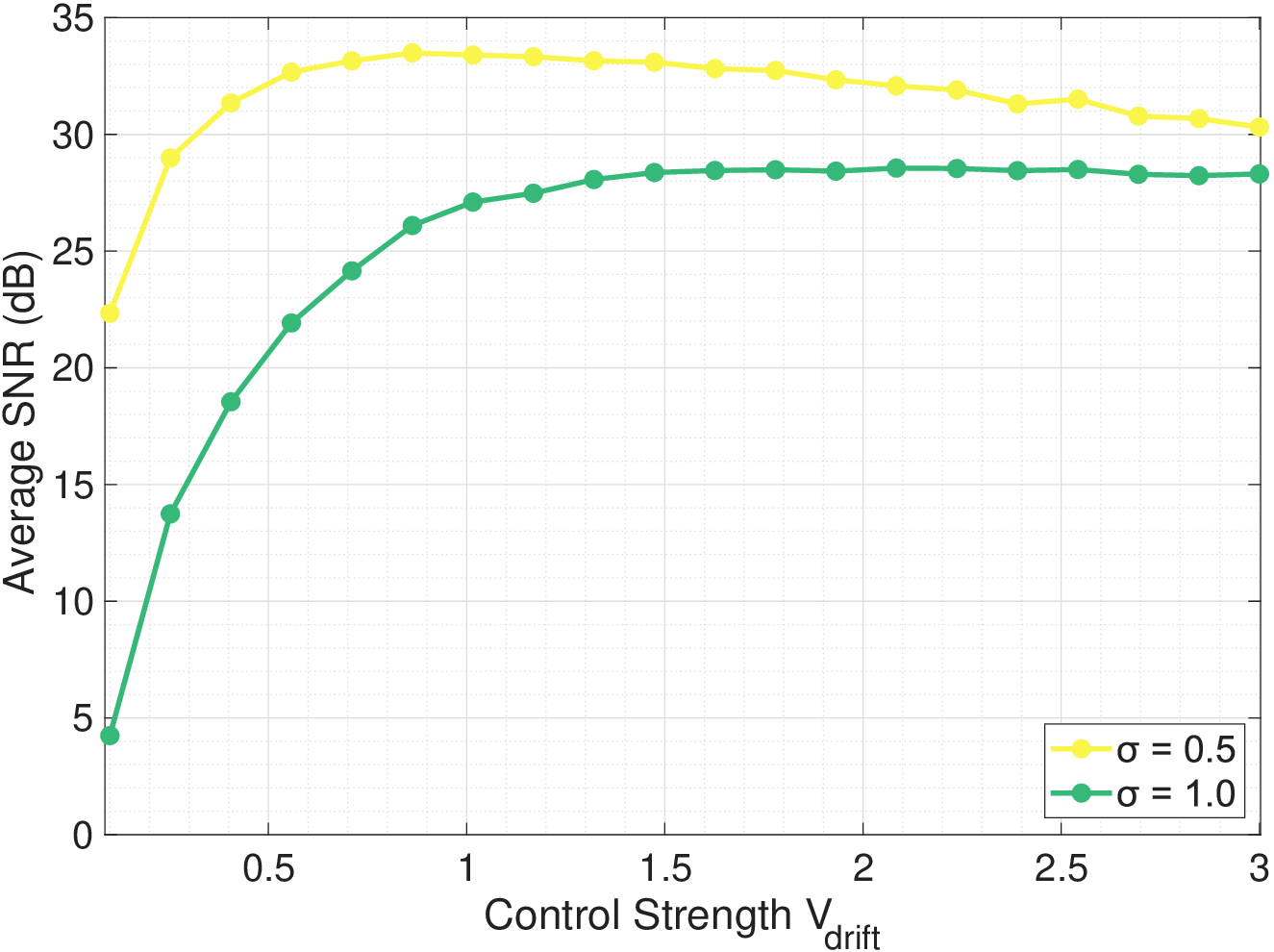}}
    \caption{Average SNR versus the control strength $V_{\text{drift}} \in [0.1, 3.0]$ for two diffusion levels: $\sigma \in \{0.5, 1.0\}$. Control strength swept across the range representing varying drift control intensity.}
    \label{se:vdrift}
    \vspace{-3mm}    
\end{figure}

Fig.~\ref{se:vdrift} examines average SNR as a function of control strength $V_{\mathrm{drift}}$ for two diffusion coefficients ($\sigma = 0.5, 1.0$). For $\sigma = 1.0$, SNR improves rapidly from 10 dB at $V_{\mathrm{drift}} = 0$ to~28 dB at $V_{\mathrm{drift}} = 1.5$, then plateaus around 27-28 dB. The $\sigma = 0.5$ case exhibits similar improvement but peaks at ~33 dB around $V_{\mathrm{drift}} = 1.0$, then shows gradual decline for $V_{\mathrm{drift}} > 1.5$. This reflects a fundamental tradeoff between spatial concentration and array diversity. At optimal points ($V_{\mathrm{drift}} \approx 1.0$-$1.5$), the equilibrium distance $\mathrm{E}[\|\mathbf{X}_n\|] = \sigma^2/V_{\mathrm{drift}} \approx 0.25$-$0.17$ m corresponds to $2.5\lambda$-$1.7\lambda$ spacing with $\lambda = 0.1$ m, allowing antennas to maintain both spatial diversity and coherent beamforming gain. However, for very strong drift ($V_{\mathrm{drift}} > 2.0$ with $\sigma = 0.5$), antennas cluster too tightly ($\mathrm{E}[\|\mathbf{X}_n\|] < 0.125$ m $\approx 1.25\lambda$), reducing effective aperture and causing phase correlation to diminish spatial degrees of freedom. Additionally, excessive control may induce rapid position fluctuations that reduce temporal coherence. The saturation validates that once $V_{\mathrm{drift}} > 2\sigma$, further increases yield diminishing returns. These results provide practical guidance: control strength should scale proportionally to environmental noise, with $V_{\mathrm{drift}} \approx 1.5\sigma$ balancing coherent gain against spatial diversity preservation.

\subsubsection{Scaling with Number of Antennas}
\begin{figure}
    \centering
    \makebox[\linewidth][c]{%
        \includegraphics[width=0.99\linewidth]{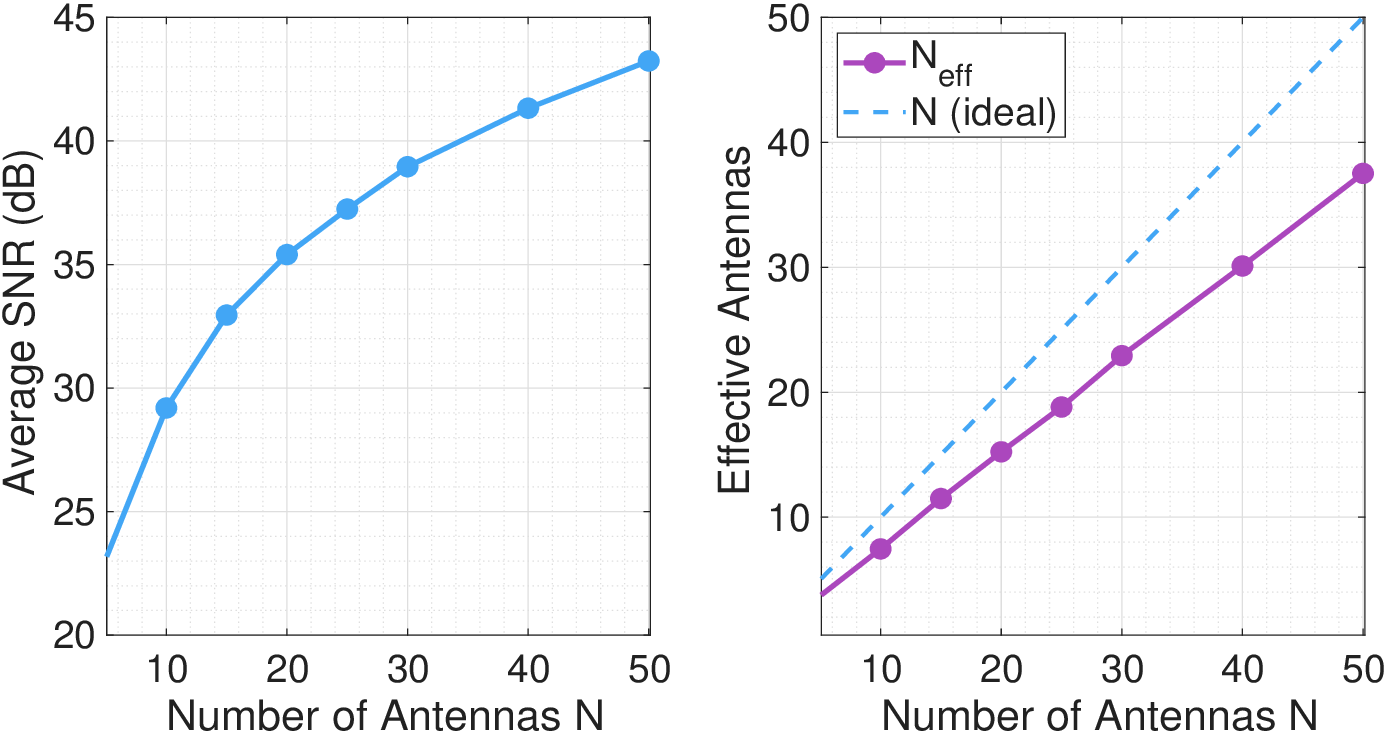}}
    \caption{System scalability analysis showing (left) average received SNR and (right) effective antenna count versus total number of antennas $N \in [5, 50]$. The left panel demonstrates logarithmic SNR scaling consistent with coherent array gain. The right panel compares the effective antenna count $N_{\text{eff}}$ (solid purple) against the ideal linear scaling $N$ (dashed blue).}
    \label{op:noante}
    \vspace{-3mm}
\end{figure}

Fig.~\ref{op:noante} demonstrates scalability properties. The left panel shows average SNR increasing from 25 dB ($N = 10$) to 43 dB ($N = 50$), validating coherent array gain preservation. The right panel reveals $N_{\text{eff}}$ scales sublinearly, reaching ~38 effective elements when $N = 50$. This efficiency ratio $N_{\text{eff}}/N \approx 0.76$ arises from stochastic positioning—at any instant, some elements occupy suboptimal locations due to diffusion. The gap between $N_{\text{eff}}$ and ideal (dashed line) quantifies the performance cost of mobility under the given control policy.
\subsection{Optimization Performance Analysis}
\begin{figure*}[htbp]
    \centering
    \makebox[\textwidth][c]{\includegraphics[width=1\linewidth, trim=20 10 10 25, clip]{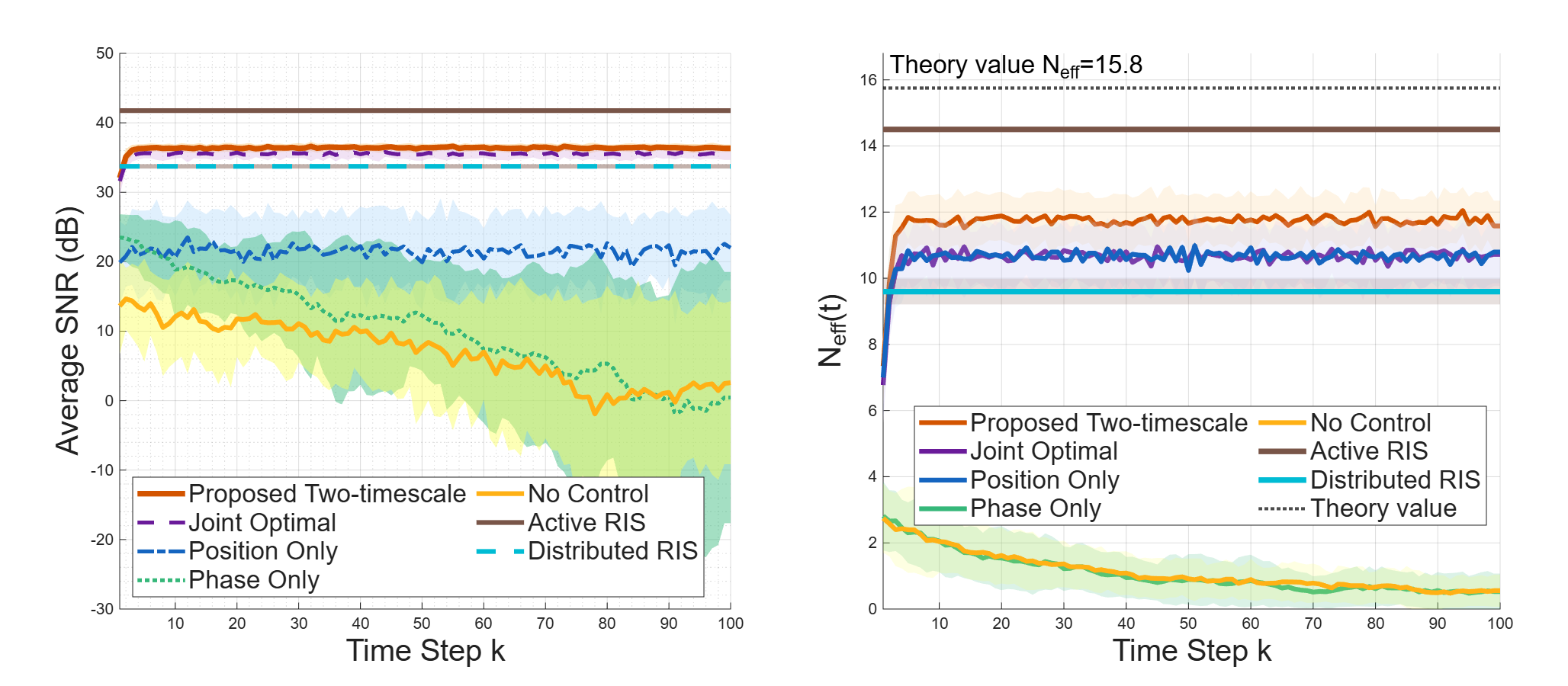}}
    \vspace{-6mm} 
    \caption{Time evolution of average received SNR (left) and effective antenna count $N_{\mathrm{eff}}$ (right), comparing seven control strategies with $V_{\mathrm{drift}} = 2.0$: (1) Proposed Two-timescale Optimization control (orange), (2) Joint Optimal trajectory and phase control (purple), (3) Position Only optimization with fixed phases (blue), (4) Phase Only optimization with random walk (green), (5) No Control (yellow), (6) Active RIS (brown), (7) Distributed RIS (cyan). Shaded regions indicate $\pm 1$ standard deviation across Monte Carlo realizations.}
    \label{op:jointop}
    \vspace{-3mm}
\end{figure*}

To contextualize the performance of our proposed MA-RIS framework, we compare five MA-RIS strategies: \textbf{Proposed Two-timescale}, \textbf{Joint Optimal}, \textbf{Position Only}, \textbf{Phase Only}, and \textbf{No Control} against two state-of-the-art fixed-position RIS architectures representing different design philosophies: \textbf{Active RIS} \cite{nguyen2024performance} and \textbf{Distributed RIS} \cite{zhang2021intelligent}. For all MA-RIS strategies, antenna positions are initialized as $\mathbf{X}_n(0) \sim \mathcal{N}(\mathbf{0}, 1.5^2 \mathbf{I}_2)$, representing elements randomly scattered around the illumination center with standard deviation 1.5~m per coordinate (comparable to the panel radius $R = 2$~m). Active RIS employs the same $N = 16$ elements arranged in a fixed $4 \times 4$ uniform grid centered at the illumination point (grid spacing optimized for maximum spatial coherence), with each element equipped with an active power amplifier providing 1.5× amplitude gain (2.25× power gain or 3.5~dB). Distributed RIS partitions $N = 16$ elements across $M = 4$ spatially separated panels, positioned at the corners of a square region with panel centers at $(\pm 0.5, \pm 0.5)$~m. The \textbf{left panel of Fig.~\ref{op:jointop}} shows the average received SNR for seven architectures:

\textbf{Proposed Two-timescale }(TSO)\textbf{ and Joint Optimal }(JO):
Both strategies drive antennas toward the illumination center and optimize phases, converging within the first 10 time steps to approximately 35--36~dB with narrow variance. At steady state, the mean radial distance tends towards the equilibrium distance: $\mathrm{E}[\|\mathbf{X}_n(t)\|] \to d_{\text{eq}} = \sigma^2/V_{\text{drift}} = 0.125$~m as $t \to \infty$. We approximate each antenna as concentrated near this equilibrium distance, giving representative antenna effectiveness $a_{\mathrm{eq}} = \exp(-d_{\mathrm{eq}}^2) = \exp(-(0.125)^2) \approx 0.984$. With $N = 16$ antennas, this yields $N_{\mathrm{eff}} \approx N \cdot a_{\mathrm{eq}} \approx 15.8$, consistent with the right panel. Coherent phase alignment then produces near-ideal array gain $\mathrm{SNR} \approx \gamma_0 N_{\mathrm{eff}}^2 \approx 35.9$~dB, matching the simulation. The key distinction is that the proposed Two-timescale method explicitly accounts for future motion dynamics and CSI overhead through backward dynamic programming (Eqs.~(\ref{eq:phase_closed_form_revised})--(\ref{eq:per_antenna_dp_revised})), achieving marginally higher and more stable performance (approximately 0.5--1~dB advantage visible in the confidence bands) compared to JO, which optimizes movement and phase using the instantaneous SNR in~(\ref{eq:inssnr}) without predictive overhead modeling.

\textbf{Position Only} (PO): With spatial control but without phase optimization, the system plateaus near 20-22 dB with $N_{\mathrm{eff}} \approx 10.5$ (right panel). In PO, phase shifts $\theta_n(t)$ are set randomly rather than aligned, resulting in incoherent combining where powers add rather than amplitudes. Assuming uniform effectiveness $\bar{a}_n \approx 10.5/16 \approx 0.656$, we get $\sum_{n=1}^{16} a_n^2 \approx 16 \times (0.656)^2 = 6.9$, giving theoretical SNR of $15 + 10\log_{10}(6.9) = 23.4$ dB. The observed 20-24 dB closely matches this prediction. This 13-15 dB gap relative to coherent strategies quantifies the critical value of phase optimization.

\textbf{Active RIS}:
The active RIS baseline achieves the highest SNR ($\approx$42~dB) through a combination of fixed optimal grid positioning and active amplification. Elements are arranged in a uniform grid centered at the illumination point with spacing optimized to maximize $N_{\mathrm{eff}} \approx 14.5$ (confirmed in right panel). Active amplification provides an additional 1.5× amplitude gain (corresponding to $(1.5)^2 = 2.25$ power gain or 3.5~dB), yielding $\mathrm{SNR} = \gamma_0 \cdot (1.5)^2 \cdot N_{\mathrm{eff}}^2 \approx 45$~dB. The flat curve behaviour reflects the absence of mobility---positions remain constant throughout, eliminating convergence transients. 

\textbf{Distributed RIS}:
Distributed RIS employs four spatially separated panels (each with $N/4 = 4$ elements) at square corners. This geometry yields $N_{\mathrm{eff}} \approx 9.5$ (right panel), lower than centralized Active RIS ($N_{\mathrm{eff}} \approx 14.4$) due to larger average distances from the illumination center. Despite fewer effective antennas, coherent phase optimization achieves~33 dB steady-state SNR—only 2-2.5 dB below TSO but 11-13 dB above PO strategy. The key distinction from PhO is that Distributed RIS maintains fixed, stable antenna positions: coherent phase control is effective when antennas occupy reasonably stable locations, but cannot compensate for unbounded spatial drift as in PhO where antennas disperse far from the illumination center.
 \begin{figure}[t]
     \centering
     \makebox[\columnwidth][c]{\includegraphics[width=1.10\linewidth, trim=20 30 25 20, clip]{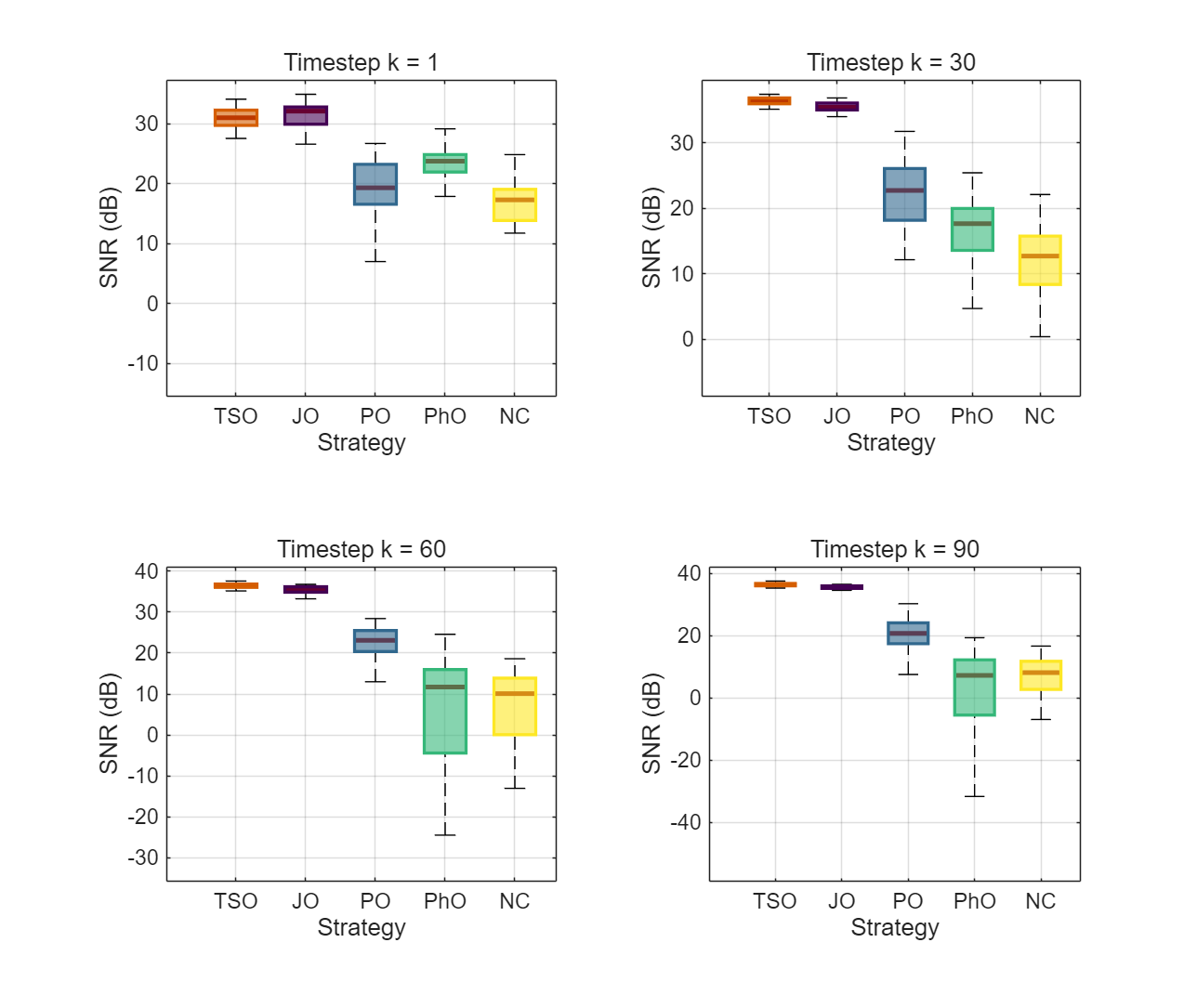}}
    \caption{SNR performance comparison at selected time steps ($k = 1, 30, 60, 90$) for five control strategies:
TSO, JO, PO, PhO, and NC, using the same color coding as Fig.~\ref{op:jointop}. Boxes show interquartile range (IQR), horizontal lines indicate medians, with $\sigma = 0.5$ and $V_{\mathrm{drift}} = 2.0$.} 
    \label{op:fill}
    \vspace{-0.2mm}
\end{figure}

\textbf{Right panel --- $N_{\mathrm{eff}}$ evolution}:
The right panel reveals that SNR hierarchy is governed by two distinct mechanisms. TSO, JO, and PO all converge to similar $N_{\mathrm{eff}} \approx 11$-$12$ (right panel), consistent with steady-state prediction in Eq.~(\ref{eq:neff}), yet their SNR differs by over 13 dB—confirming $N_{\mathrm{eff}}$ captures only spatial positioning while phase alignment provides coherent combining gain. The horizontal dashed line represents theoretical steady-state $N_{\mathrm{eff}}$, calculated by evaluating Eq.~(\ref{eq:neff}) at equilibrium distance from Eq.~(\ref{e25}). For $V_{\mathrm{drift}} = 2.0$, the predicted value is $N \cdot \exp(-d_{\text{eq}}^2) = 16 \times \exp(-0.016) \approx 15.8$. Position-controlled strategies (TSO, JO, PO) converge to stable $N_{\mathrm{eff}}$ (~10.5-12), while fixed deployments (Active RIS, Distributed RIS) remain constant at ~14.5 and 9.5 respectively. Only uncontrolled strategies (PhO, NC) degrade continuously to near zero, confirming that spatial stability—through either controlled drift or fixed positioning—is essential to counteract environmental diffusion.

Fig.~\ref{op:fill} statistically validates these trends across 40 realizations at $k = 1, 30, 60, 90$. At $k = 1$, all strategies show positive SNR (TSO/JO median ~31 dB; PO ~20 dB; PhO ~24 dB; NC ~17 dB). By $k = 30$, stratification is clear: TSO/JO tighten near 35-36 dB (IQR $<$ 2 dB), PO stabilizes at ~20 dB, while PhO and NC degrade with increasing spread. At $k = 90$, TSO/JO maintain median ~37-38 dB with IQR $<$ 2 dB, and PO holds at ~22 dB. In contrast, PhO and NC show medians near 5-10 dB with wider spread—PhO extends below -5 dB versus NC reaching 2 dB. PhO's larger variance occurs because coherent combining with optimized phases is highly sensitive to antenna positioning: when antennas are randomly dispersed, coherent gain varies drastically across realizations—from constructive interference when several antennas happen to be moderately positioned to destructive interference when most drift far apart. Three key findings emerge: (i) TSO and JO provide superior performance and remarkable stability, essential for QoS guarantees; (ii) PO achieves moderate but stable performance; (iii) strategies lacking position control exhibit increasingly unpredictable performance. By $k=90$, PhO and NC show IQR exceeding 15-20 dB with whiskers to -20 dB, indicating substantial dispersion. This heterogeneity creates highly skewed SNR distributions that degrade system reliability.
 \begin{figure}[t]
     \centering
     \makebox[\columnwidth][c]{\includegraphics[width=1.0\linewidth]{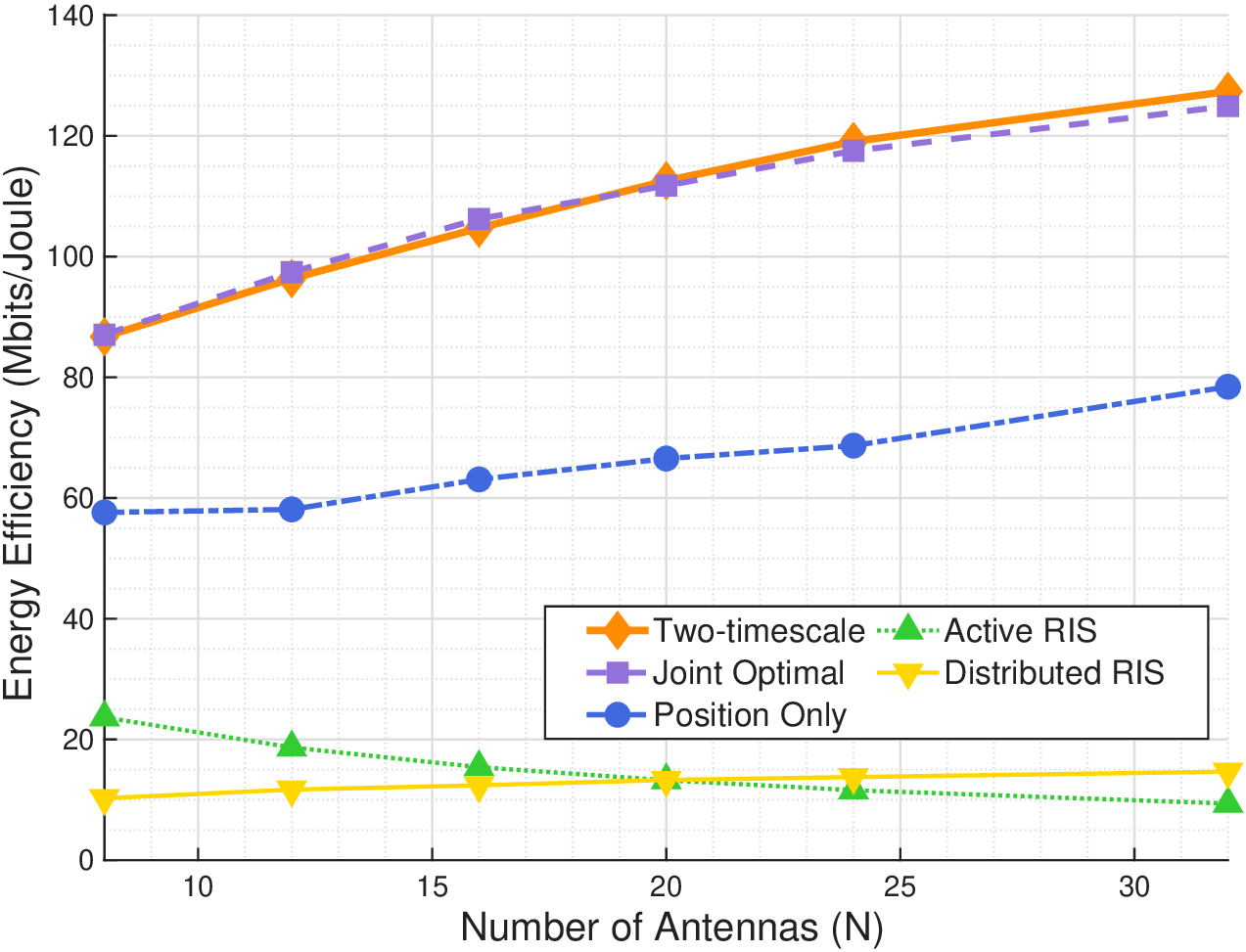}}
    \caption{EE versus number of antennas over five architectures compared with $\gamma_0 = 15$ dB, $\sigma = 0.5$, and $V_{\mathrm{drift}} = 2.0$.}
\label{fig:ee_vs_n} 
\vspace{-2mm}
\end{figure}
 \begin{figure}[t]
     \centering
     \makebox[\columnwidth][c]{\includegraphics[width=1.0\linewidth]{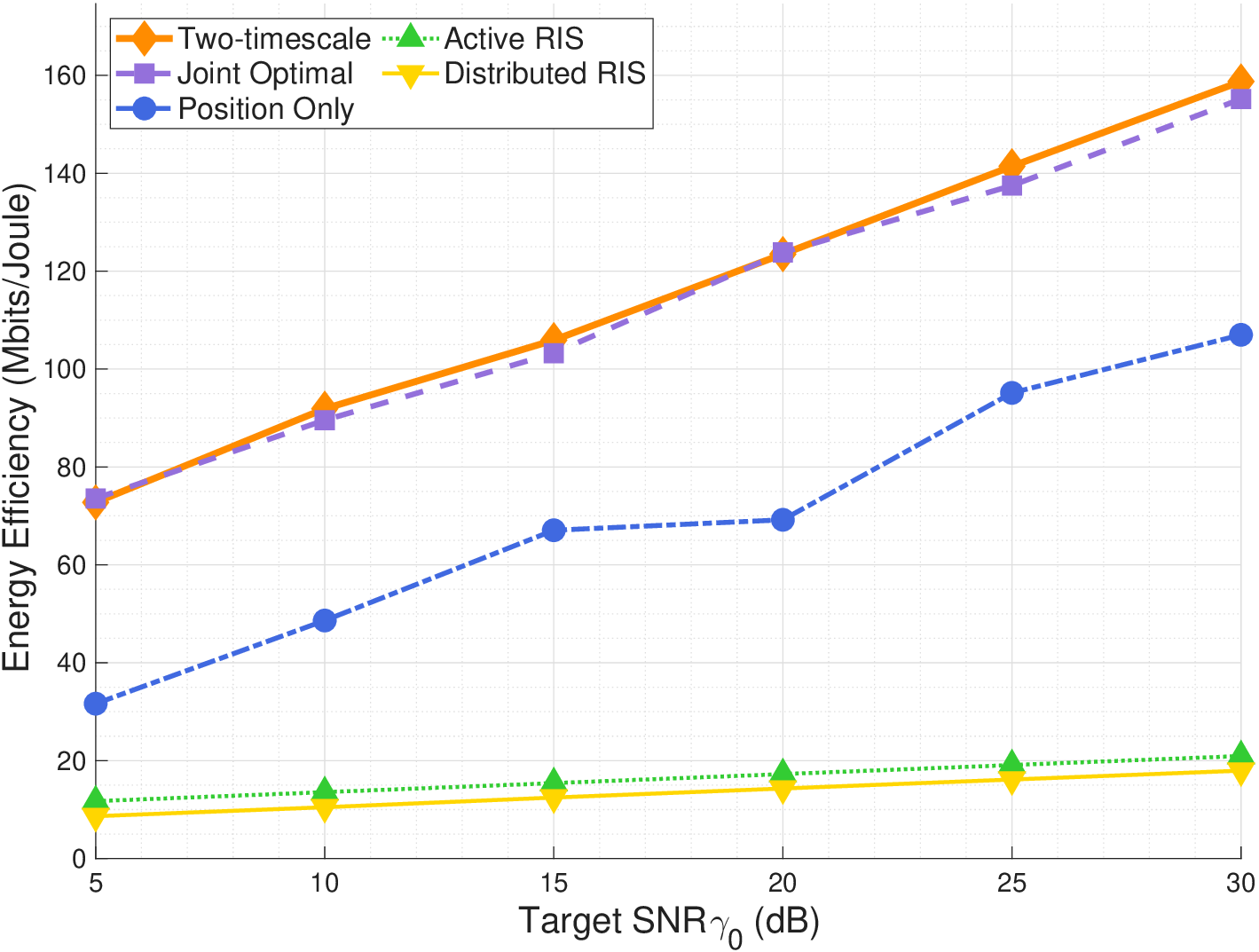}}
    \caption{Energy efficiency versus target SNR over five architectures compared with $\sigma = 0.5$, and $V_{\mathrm{drift}} = 2.0$.}
\label{fig:ee_vs_snr}
\vspace{-2mm}
\end{figure}
\subsection{Energy Efficiency Comparison}
Fig.~\ref{fig:ee_vs_n} shows EE versus number of antennas at $\gamma_0 = 15$ dB. PhO and NC are omitted due to negligible data rates from severe SNR degradation. The proposed TSO achieves 87-127 Mbits/J as $N$ increases from 8 to 32, maintaining 2-3\% advantage over JO with favorable scaling. Active RIS degrades from 25 to 8 Mbits/J as $N$ increases because amplifier power scales linearly while SNR gains saturate. At $N = 32$, MA-RIS based TSO delivers 16 times the EE of Active RIS (127 vs. 8 Mbits/J); at $N = 16$, this advantage is 7 times (105 vs. 15 Mbits/J). PO achieves 58-78 Mbits/J, demonstrating spatial control alone provides substantial EE but sacrifices 30-35\% relative to coherent strategies. Distributed RIS performs poorly (10-15 Mbits/J) due to reduced spatial effectiveness.
Fig.~\ref{fig:ee_vs_snr} examines EE versus target SNR $\gamma_0$ at fixed $N = 16$. As channel quality improves from 5 to 30~dB, the proposed Two-timescale strategy maintains higher efficiency than Active RIS across all conditions: at $\gamma_0 = 5$~dB, MA-RIS achieves 75~Mbits/J versus Active RIS's 12~Mbits/J; at $\gamma_0 = 30$~dB, the advantage persists at 160~Mbits/J versus 20~Mbits/J. This consistent gap demonstrates that MA-RIS's EE benefit is robust to channel variability and does not rely on favorable propagation conditions. The TSO and JO curves remain nearly overlapping (2--3\% difference) across the entire SNR range, confirming that overhead-aware predictive optimization provides measurable but modest refinement and that position control is the dominant performance factor. PO tracks 30\% below coherent MA-RIS strategies across all $\gamma_0$ values, quantifying the fixed coherent combining advantage regardless of base channel quality. For practical deployments, MA-RIS enables longer-hour operation on a battery-limited devices versus Active RIS, positioning it as the preferred architecture for energy-constrained scenarios where moderate SNR with minimal power outweighs peak performance with excessive consumption.

\section{Conclusion}
This paper presented a stochastic control framework for movable antenna-enhanced reconfigurable intelligent surfaces (MA-RIS) that treats antenna mobility as a controllable degree of freedom to combat static deep fades. By modeling antenna dynamics as stochastic differential equations, we derived the steady-state equilibrium $\mathrm{E}[\|\mathbf{X}_n(t)\|] = \sigma^2/V_{\mathrm{drift}}$ and showed that movements exceeding $0.4\lambda$ effectively decorrelate spatial fading. Our analytical characterization of outage probability and effective antenna count provides practical design guidelines, achieving approximately 76\% antenna utilization under stochastic control. 
We proposed a Two-timescale overhead-aware optimization framework decomposing joint movement-phase control into slow predictive position optimization via backward dynamic programming and fast phase alignment. By incorporating CSI and mobility overhead, the method achieves 35-36 dB steady-state SNR with remarkable stability (IQR $<$ 2 dB)—13-15 dB gain over position-only control and over 30 dB improvement relative to uncontrolled baselines. Despite about 6 dB lower SNR than Active RIS, MA-RIS achieves 7-13 times higher EE through minimal power consumption, demonstrating the superiority of moderate SNR with low power, over peak performance with excessive consumption. Future directions include Doppler-aware optimization for mobility, multi-user coordination, deep reinforcement learning for scalability, hardware validation, practical CSI modeling, and quantized positioning control.


\bibliographystyle{unsrt}
\bibliography{Citation.bib}

\end{document}